\documentclass[preprintnumbers,11pt,prd,showpacs]{revtex4}

\usepackage{graphicx}

\usepackage{amssymb}

\begin{document} 
\preprint{BU-HEPP-08-07}

\title{Thomas-Fermi Statistical Models of Finite Quark Matter}
\thanks{Partialty supported by the Baylor University Quantum Optics Initiative.}

\author{Walter Wilcox}
\thanks{walter\_wilcox@baylor.edu}

\affiliation{Department of Physics, Baylor University, Waco, TX 76798-7316}

\begin{abstract}
I introduce and discuss models of finite quark matter using the formalism of the
Thomas-Fermi statistical model. Similar to bag models, a vacuum energy density term is introduced to model long distance confinement, but the model produces bound states from the residual color Coulomb attraction even in the absence of such a term. I discuss three baryonic applications: an equal mass nonrelativistic model with and without volume pressure, the ultra-relativistic limit confined by volume pressure, and a color-flavor locking massless model. These models may be extended to multi-meson and other mixed hadronic states. Hopefully, it can help lead to a better understanding of the phenomenology of high multi-quark states in preparation for more detailed lattice QCD calculations.
\end{abstract}

\pacs{12.39.-x,12.39.Mk,12.40.Ee,21.65.Qr}

\maketitle
\newpage

\section{Introduction}

A long standing question in the theory of quark matter asks: where are the mesons, baryons or mixed states with more than the usual two or three quarks? It is likely from the theory, quantum chromodynamics (QCD), that high multi-quark states exist. A number of \lq\lq exotic" hadronic objects have been postulated. There has been speculation concerning hypothetical strange quark states\cite{witten,farhi,sreview} for many years, partly motivated by astrophysical considerations involving the so-called GKZ cutoff for suppression of high-energy cosmic rays\cite{madsen1}. It hasn't been clear until recently that the GZK cutoff mechanism was producing the expected suppression of events at energies above $\sim 10^{20}$ GeV\cite{recent}. 

Experimental evidence for a four quark one antiquark state has been much discussed\cite{rpp} since the initial report in 2003\cite{penta}, and there is a very recent report of a double meson four quark state from the Belle collaboration\cite{4q}. The best way to investigate such theoretical questions is with the set of methods known as lattice QCD. Although it is in the initial stages of being applied to larger nuclear systems lattice QCD is, however, still limited to small volumes and quark numbers. There is a need for models which can help lead expensive lattice QCD calculations in the right direction in the search for high quark states. One such treatment is the MIT bag model, which has been applied to the problem of modeling hadronic bound states with many quarks\cite{farhi,madsen1,madsen2}. The Nambu-Jona-Lasinio model has also been used in such analyses\cite{kiri}.

I would like to point out here that the Thomas-Fermi (TF) treatment of particles as a gas at zero temperature\cite{TF,spruch} provides a versatile and adaptable set of models which can be used to study such questions. It has been quite successful in treating atomic systems, viewing those as a bound collection of interacting gas particles, even for low atomic numbers. One of its attractive features is the exact treatment it affords to the Coulomb interactions. It can be understood as a semi-classical approach which correctly incorporates fermion statistics for a large number of particles but is not fully quantum mechanical. It has traditionally been used to model ground state atomic systems (it has also been applied to nuclear and condensed matter problems), but can also be applied to quark systems. We will find that this is a natural extension of the method.

The following model combines the TF teatment of quarks with a bag model assumption to model confinement. It removes the central nuclear force present in the traditional TF atomic model and replaces it with a collective residual color Coulombic interaction which provides binding even in the absence of a volume term. Although it is related to bag models, the TF quark model has a different starting point. Bag models represent exact solutions to free fermions placed in a spherical cavity. In the TF model, one deals not with individual modes but with a single, collective, spherically symmetric density of states which represents a gas of $3A$ (\lq\lq A" is baryon number) interacting fermions at zero temperature. 
The spherical symmetry assumption is natural for a gas of particles in their ground states and is {\it a prosteriori} justified by bag model fits, with reasonable phenomenology. The quarks interact via their residual color Coulombic interactions. (The word \lq\lq residual" is meant to denote the non-primary, non-confining color Coulombic part of the interaction. This interaction scales like $\sim 1/A$ in a color singlet, and is thus residual in a large $A$ sense also.) In contrast, bag models assume no interactions to lowest order and the color interactions are put in perturbatively. In other words, the advantage for TF models is the nonperturbative inclusion of the Coulombic interactions; the disadvantage that one is no longer working with exact solutions of the Dirac equation but assumes a statistical treatment. Nevertheless, one would expect that the TF quark model would become increasingly accurate as the number of constituents is increased, that is, as a statistical treatment becomes more justified. Note we are talking only about ground states in this context, as the TF model is incapable of describing excited atomic states. The main usefulness of the model will be in seeing systematic trends and effects in ground states as the parameters of the model are varied and helping to identify likely stable bound states.

The different starting point for the TF quark models results in a very different consequence relative to bag models: the residual color Coulombic interactions produce a bound state in the massive version of the model. The attractive collective potential is shown in Fig.~\ref{NRpot} below for the nonrelativistic single flavor case. However, these Coulombic bound states are very different from normal hadronic states. Their energies scale like $A^{1/3}$ rather than $A$. The natural interpretation of these systems is that they are bound states in a zero temperature version of the quark-gluon (\lq\lq deconfined") phase of QCD. 

In order to describe normal hadrons, it is also natural to assume a phase transition to a confined state. This is implemented by the introduction of the bag energy, $BV$ ($V$ is the volume), which produces an external pressure and results in a surface discontinuity in the particle density function. Then, as a result of the total energy minimization, one finds these energies are proportional to $A$ for $A>>1$. The nonrelativistic model has a total energy as a function of $A$  given by Eq.(\ref{etot}). This equation is similar to the liquid-drop nuclear model binding energy equation. Deviations from the $\sim A$ behavior should tell us where to expect greater or lesser binding. The (massless) relativistic model's energy equation, Eq.(\ref{relenergy}), also scales like $A$ for $A>>1$.

In the following section I will first work out the Coulombic couplings associated with a color singlet gas of  quarks whose charges are not fixed. The interactions in the TF quark model are modeled on the basis of classical QCD, as explained for example in Ref.\cite{SandW}. As mentioned above, the effective couplings, which incorporate color interaction probability factors deduced from the classical theory, decrease like $1/A$ for large values of $A$. The energy functional will then be formulated in Section III where I will show that the energy equations can be formally integrated and related to endpoint values and derivatives of the TF function. I consider a nonrelativistic form of the model with $N_f$ mass degenerate quarks in Section IV. A stable state is formed, which, as discussed above, is interpreted as a hadronic bound state in the zero temperature quark-gluon phase. In Section V I examine the effects of turning on the external pressure, and we will see that it produces a surface discontinuity in the TF functions. An explicit expression for the system energy will be derived. I will briefly consider the general form of the TF relativistic quark equation in Section VI before solving the specific case of  $N_f$ massless quarks. This system is subjected to external pressure and an expression for the system energy is derived. To show the versatility of the model and point to future applications, I also consider aspects of a color-flavor locking type interaction in Section VII, which includes a Cooper pairing term and massive gluons. I close with a summary and some general comments about directions of future work in Section VIII.

During final editing of this paper, the author became aware of the paper\cite{dixon}, which uses an approximate TF expression for the fermion partition function to evaluate the equation of state for a gas of massless quarks interacting with $SU(2)$ or $SU(3)$ color gluons. I will comment more on this paper and its relationship with the present  model in Section VI. See also Refs.\cite{spence1,spence2}, which applies a TF approximation to the evaluation of the lattice QCD fermion partition function.

\section{Residual Coulombic Couplings}

Quark-quark and quark-antiquark $SU(3)$ color interactions at the classical level can be understood from their \lq\lq charges" in isospin/hypercharge space, usually designed as the \lq\lq 3" and \lq\lq 8" axes\cite{SandW}. Quarks can be understood to have strong interaction charges with magnitude $\frac{4}{3}g^2$, where $g$ is the strong interaction coupling constant, located at the corners of an equilateral triangle with vertices at $(-1,\frac{1}{\sqrt{3}})g$, $(1,\frac{1}{\sqrt{3}})g$, and $(0,-\frac{2}{\sqrt{3}})g$ in $(3,8) $ space notation. Likewise, antiquarks can be understood to exist at the points  $(-1,-\frac{1}{\sqrt{3}})g$, $(1,-\frac{1}{\sqrt{3}})g$, and $(0,\frac{2}{\sqrt{3}})g$. In this section I will treat only multi-baryon states and will assume that all bound states have zero net color. Introducing the total charge as
\begin{equation}
\vec{Q}=\sum_{i=1}^{3A}\vec{q}_i,
\end{equation}
where the sum is over the $3A$ quarks, we have for its square,
\begin{equation}
\vec{Q}^2=\sum_{i=1}^{3A}\vec{q}\,_i^2 +2 \sum_{i\ne j}\vec{q}_i\cdot\vec{q}_j.
\end{equation}
The magnitude of any single term is
\begin{equation}
\vec{q}\,_i^2=\frac{4}{3}g^2.
\end{equation}
For $i\ne j$ and $3A$ quarks, one encounters the same color (repulsive) and different color (attractive) interactions
\begin{equation}
\vec{q}_i\cdot\vec{q}_j=\left\{ \begin{array}{l} \frac{4}{3}g^2,  3A(A-1)/2 \,\,{\rm times} \\
-\frac{2}{3}g^2,  3A^2 \,\,{\rm times} \end{array}\right.
\end{equation}
respectively. This leads to
\begin{equation}
\vec{Q}^2=\frac{4}{3}g^2(3A + 2 (\frac{3A(A-1)}{2}-\frac{1}{2}3A^2))=0,
\end{equation}
as it should for an overall color singlet. The average repulsive coupling between quarks with the same color is
\begin{equation}
\left(\frac{\frac{3A(A-1)}{2}}{\frac{3A(3A-1)}{2}}\right)\frac{4}{3}g^2=\left(\frac{A-1}{3A-1}\right)\frac{4}{3}g^2,\label{repul}
\end{equation}
and the average attractive coupling for different colored quarks in a baryon is
\begin{equation}
\left(\frac{3A^2}{\frac{3A(3A-1)}{2}}\right)(-\frac{2}{3}g^2)=\left(\frac{2A}{3A-1}\right)(-\frac{2}{3}g^2).\label{attract}
\end{equation}

The TF quark model replaces the sum over particle number in particle interaction models with a sum over the density of state particle properties. The interaction strengths are taken from the classical theory, but it is necessary to weight these by the probabilities of the various interactions in the color sector, which we assume are flavor independent. We need a connection between particle number and probability. The natural assumption is that these interaction probabilities are proportional to the number of particle interaction terms. Thus, we assume that the average repulsive (same color) or attractive (different color) interaction couplings, Eqs.(\ref{repul}) and (\ref{attract}), equal to $\frac{4}{3}g^2\sum_{i}P_{ii}$ and $-\frac{2}{3}g^2\sum_{i\ne j}P_{ij}$, respectively. In addition, we assume that all colors in a color singlet contribute equally to these sums. We therefore have
\begin{equation}
P_{ii}=\frac{1}{3}\left( \frac{A-1}{3A-1} \right),
\end{equation}
for self-color interactions and 
\begin{equation}
P_{ij}=\frac{1}{3}\left( \frac{A}{3A-1} \right) (i\ne j),
\end{equation}
for different colors. These sum to one as they should:
\begin{equation}
\sum_{i,j} P_{ij}=1.
\end{equation}
Using these, we can now construct the nonrelativistic interaction energy, $E$, from the kinetic and color Coulomb parts of the interaction to form the nonrelativistic system energy. This will be done in Section III. The overall \lq\lq residual" coupling, the sum of (\ref{repul}) and (\ref{attract}), is
\begin{equation}
\left(\frac{(A-1)-A}{3A-1}\right)\frac{4}{3}g^2=-\frac{\frac{4}{3}g^2}{3A-1},\label{coupling}
\end{equation}
which is just the negative of the charge of a single quark divided by the number of remaining quarks.

\section{System Energy and Equations}

Assuming flavor and color number, $n^I_i(p_{F})$, and non-interacting quark kinetic energy, ${\cal E}^I_i(p_{F})$, densities, we have that these are related to the Fermi momenta, $(p_F)^I_i$ by
\begin{equation}
n^I_i(p_{F})=2\int^{p_F}  \frac{d^3p^I_i}{(2\pi \hbar)^3}=\frac{((p_F)^I_i)^3}{3\pi^2\hbar^3}\label{np_F},\label{n}
\end{equation}
and
\begin{equation}
{\cal E}^I_i(p_{F})=2\int^{p_F}  \frac{d^3p^I_i}{(2\pi \hbar)^3}\frac{(p^I_i)^2}{2m}=\frac{(3\pi^2\hbar^3n^I_i(p_F))^{5/3}}{10\pi^2\hbar^3m},\label{E}
\end{equation}
where the \lq\lq $I$" superscript stands for flavor and the \lq\lq $i$" subscript stands for color. According to the assumptions of the TF model, all quantities are derived from the particle densities, $n_i^I$. Thus, according to Eqs.(\ref{n}) and (\ref{E}), the kinetic energy density is proportional to $(n_i^I)^{5/3}$. The model assumes that this relationship holds at each point in space, $r$, and the ground state of the system is obtained by minimizing the energy functional. I assume the flavor-summed spatial normalization is given by
\begin{equation}
\sum_{I}\int d^3r\, n^{I}_i(r)=A.
\end{equation}
The number of quarks with flavor $I$ is designated $N^I$. The color-summed number is therefore
\begin{equation}
\sum_{i}\int d^3r\, n^{I}_i(r)=N^I,
\end{equation}
and total number is
\begin{equation}
\sum_I N^I = 3A.
\end{equation}
For convenience, I will also introduce the single-particle normalized density
\begin{equation}
{\hat n}^{I}_i\equiv \frac{3n_i^I}{N^I}.
\end{equation}
This form of the density will be helpful in correctly normalizing the TF quark-quark interaction energy when continuum sources are used.

I use the the color interaction probabilities from Section II to construct the nonrelativistic system energy, $E$. I use the notation \lq\lq $i<j$" on the color sums below to avoid double counting. The minus signs and factors of 1/2 in front of the color sums involving $P_{ij}$ are building in the interaction strengths from Eqs.(\ref{repul}) and (\ref{attract}). I double count on the flavor sum, $I\ne J$, and compensate with a factor of two. Using the single particle densities, ${\hat n}^I_i(r)$, I normalize to the number of terms one has from the discrete form of the same-flavor interaction, $N^I(N^I-1)/2$, and the number of terms in the different flavor interaction, $N^IN^J$. (I will check on the energy normalizations in the next section.) One has
\begin{eqnarray}
\nonumber E=T+U=\sum_{i,I} \int^{r_{max}}d^3r\frac{(\pi^2\hbar^3N^I{\hat n}^I_i(r))^{5/3}}{10\pi^2\hbar^3m^I}\quad\quad\quad\quad\quad\quad\\ 
\nonumber +\frac{4}{3}g^2 \sum_I \frac{N^I(N^I-1)}{2}
\int^{r_{max}}\int^{r_{max}}\frac{d^3r\,d^3r'}{|{\vec r}-{\vec r}\,'|}
\left(\sum_i P_{ii}{\hat n}^I_i(r){\hat n}^I_i(r') \right. \\
\nonumber \left. -\frac{1}{2}\sum_{i< j} P_{ij}{\hat n}^I_i(r){\hat n}^I_j(r')\right)\\
\nonumber +\frac{4}{3}g^2\sum_{I\ne J}\frac{N^IN^J}{2}\int^{r_{max}}\int^{r_{max}}\frac{d^3r\,d^3r'}
{|{\vec r}-{\vec r}\,'|}\left(\sum_i P_{ii}{\hat n}^I_i(r){\hat n}^J_i(r') \right. \\
\left. -\frac{1}{2}\sum_{i< j} P_{ij}{\hat n}^I_i(r){\hat n}^J_j(r')\right).
\end{eqnarray}
I have assumed that the radius of the objects introduced are finite, and indicate this with the \lq\lq $r_{max}$" notation on the spatial integrals. I now do two things: switch to normalization $n^I_i$ and assume equal Fermi color momenta, $n^I\equiv n_1^I=n_2^I=n_3^I$, for each $I$. Doing the color sums, we have
\begin{eqnarray}
\nonumber E&=&\sum_{I} \int^{r_{max}}d^3r\frac{3(3\pi^2\hbar^3n^I(r))^{5/3}}{10\pi^2\hbar^3m^I}\\ 
\nonumber & - &\frac{ 9\times\frac{4}{3}g^2}{2(3A-1)}
\sum_I \frac{N^I-1}{N^I} \int^{r_{max}}\int^{r_{max}}d^3r\,d^3r'\frac{n^I(r)n^I(r')}{|{\vec r}-{\vec r}\,'|} \\
\nonumber &-&\frac{ 9\times\frac{4}{3}g^2}{2(3A-1)}\sum_{I\ne J}\int^{r_{max}}\int^{r_{max}}d^3r\,d^3r'\frac{n^I(r)n^J(r')}{|{\vec r}-{\vec r}\,'|} \\ 
&+&\sum_I \lambda^I \left(3\int^{r_{max}} d^3r\,n^I(r)-N^I\right),
\end{eqnarray}
where I have introduced Lagrange multipliers, $\lambda^I$, associated with the constraint
\begin{eqnarray}
\int^{r_{max}} d^3r\,n^I(r)=N^I/3.\label{constraint}
\end{eqnarray}
I now do the variation of the density, $\delta n^I(r)$, in $E$.  (The steps of setting $n_1^I=n_2^I=n_3^I$ and doing the variation are interchangeable.) The result may be written
\begin{eqnarray}
\frac{(p_F^I)^2} {2m^I}=-\lambda^I +
\nonumber \frac{3\times\frac{4}{3}g^2}{(3A-1)}\left(\frac{N^I-1}{N^I} \int^{r_{max}}\!\!d^3r' 
\frac{n^I(r')}{|{\vec r}-{\vec r}\,'|}+\sum_{J\ne I}\int^{r_{max}}\!\!d^3r' \frac{n^J(r')}{|{\vec r}-{\vec r}\,'|} \right),\\ \label{Eequation}
\end{eqnarray}
with $p_F^I$ given in terms of $n^I$ by Eq.(\ref{np_F}). Notice the residual coupling, Eq.(\ref{coupling}), as an overall factor in front of the potential term on the right hand side. I introduce
\begin{equation}
f^I(r) \equiv   \frac{ra}{2\times\frac{4}{3}\alpha_s}(3\pi^2 n^I(r))^{2/3}\label{FinN},
\end{equation}
for the TF spatial functions, where
\begin{equation}
a\equiv \frac{\hbar}{m^1c},
\end{equation}
is the reduced Compton wavelength and
\begin{equation}
\alpha_s\equiv \frac{g^2}{\hbar c},
\end{equation}
defines the strong coupling constant. I choose the scale \lq\lq $a$" to be associated with the lightest quark mass, which I have designated as $m^1$. Since the masses, $m^I$, are not necessarily equal, I will choose to order, for example, $m^1\le m^2\le m^3$. Assuming spherical symmetry
\begin{equation}
 \int^{r_{max}}\!d^3r' \frac{n^I(r')}{|{\vec r}-{\vec r}\,'|} =4\pi \left[  \int_0^r dr'r'\,^2 \frac{n^I(r')}{r}+ \int_r^{r_{max}}\!dr' r'\,^2 \frac{n^I(r')}{r'}  \right],
\end{equation}
one now has the quark model TF integral equations,
\begin{eqnarray}
\nonumber \alpha^If^I(r) &=&  -\frac{\lambda^I r}{\frac{4}{3}g^2} +\frac{4(\frac{4}{3}\alpha_s)^{3/2}}{\pi(3A-1)}\left[  \frac{N^I-1}{N^I} \left( \int_0^r dr'r'\,^2 \left(\frac{2f^I(r')}{ar'}\right)^{3/2}\right.\right. \\
\nonumber &+& \left. \left. r \int_r^{r_{max}}\!dr' r' \left( \frac{2f^I(r')}{ar'} \right)^{3/2}\right) +
\sum_{J\ne I} \left( \int_0^r dr'r'\,^2 \left(\frac{2f^J(r')}{ar'}\right)^{3/2}\right.\right. \\
&+& \left. \left. r \int_r^{r_{max}}\!dr' r' \left( \frac{2f^J(r')}{ar'} \right)^{3/2}\right)  \right],\label{TFintegral}
\end{eqnarray}
where $\alpha^I\equiv m^1/m^I$. It is convenient to introduce a dimensionless distance, $x$, given by
\begin{equation}
r = Rx\label{vtrans},
\end{equation}
where the physical distance $R$ is given by
\begin{equation}
R \equiv \left(\frac{a}{2\times\frac{4}{3}\alpha_s}\right)\left[\frac{3\pi A}{4} \right]^{2/3},\label{peq}
\end{equation}
We now have
\begin{eqnarray}
\nonumber \alpha^If^I(x) &=&  -\frac{\lambda^I Rx}{\frac{4}{3}g^2} +\frac{A}{A-\frac{1}{3}}  \left[  \frac{N^I-1}{N^I} \left( \int_0^x dx'x'\,^2 \left(\frac{f^I(x')}{x'}\right)^{3/2}\right.\right. \\
\nonumber &+& \left. \left. x \int_x^{x_{max}}\!dx' x' \left( \frac{f^I(x')}{x'} \right)^{3/2}\right) +
\sum_{J\ne I} \left( \int_0^x dx'x'\,^2 \left(\frac{f^J(x')}{x'}\right)^{3/2}\right.\right. \\
&+& \left. \left. x \int_x^{x_{max}}\!dx' x' \left( \frac{f^J(x')}{x'} \right)^{3/2}\right)  \right].
\end{eqnarray}
The first derivative of this is
\begin{eqnarray}
\nonumber \alpha^I\frac{df^I(x)}{dx} =  -\frac{\lambda^I R}{\frac{4}{3}g^2} +\frac{A}{A-\frac{1}{3}} &&\!\!\!\!\left[  \frac{N^I-1}{N^I} \left( \int_x^{x_{max}}\!dx' x' \left( \frac{f^I(x')}{x'} \right)^{3/2}\right) \right. \\
&+&\sum_{J\ne I}  \left. \int_x^{x_{max}}\!dx' x' \left( \frac{f^J(x')}{x'} \right)^{3/2} \right],
\end{eqnarray}
and the second derivative is
\begin{eqnarray}
\alpha^I\frac{d^2f^I(x)}{dx^2} =  -\frac{A}{A-\frac{1}{3}} \frac{1}{\sqrt{x}}&& \!\!\!\! \left[  \frac{N^I-1}{N^I} \left( f^I(x) \right)^{3/2} 
+ \sum_{J\ne I}\left( f^J(x) \right)^{3/2} \right].
\end{eqnarray}
The normalization integral, Eq.(\ref{constraint}), reads
\begin{equation}
\int_0^{x_{max}} dx \sqrt{x}(f^I(x))^{3/2}=\frac{N^I}{3A}.
\end{equation}

\section{Nonrelativistic $N_f$-Flavor Model}

I will solve the TF inegral equations, (\ref{TFintegral}), for the convenient case of $N_f$ mass-degenerate quark flavors with equal numbers, $N^I$, even though it may not be realistic. One would expect increased stability for larger $N_f$ from Fermi statistics and this will be examined below. We have 
\begin{equation}
N^I=\frac{3A}{N_f}.
\end{equation}
Defining $n(r)\equiv n^I(r)$, $f(x)\equiv f^I(x)$, and $\lambda \equiv \lambda^I$, where from Eq.(\ref{FinN})
\begin{equation}
n(r)=\frac{1}{3\pi^2}\left( \frac{2\times \frac{4}{3}\alpha_s}{a} \right)^3 \left( \frac{4}{3\pi A}\right)\left(\frac{f}{x} \right)^{3/2},
\end{equation}
we have
\begin{equation}
\int^{r_{max}} \!\!d^3r\,n(r) = \frac{A}{N_f}.
\label{singlenorm}
\end{equation}
In terms of the TF function, $f(x)$, one has
\begin{equation}
\int_0^{x_{max}} \!\!dx \sqrt{x}(f(x))^{3/2}=\frac{1}{N_f}.\label{norm1}
\end{equation}
The $N_f$-degenerate TF integral equation becomes
\begin{eqnarray}
f(x) =   -\frac{\lambda Rx}{\frac{4}{3}g^2} + N_f\left[ \int_0^x dx'x'\,^2\left(\frac{f(x')}{x'}\right)^{3/2}
+  x \int_x^{x_{max}}\!\!dx' x' \left( \frac{f(x')}{x'} \right)^{3/2}\right].\label{f}
\end{eqnarray}
The first and second derivatives are
\begin{eqnarray}
\frac{df(x)}{dx} =  -\frac{\lambda R}{\frac{4}{3}g^2} +N_f\int_x^{x_{max}}\!\!dx' x' \left( \frac{f(x')}{x'} \right)^{3/2},\label{df}
\end{eqnarray}
and
\begin{eqnarray}
\frac{d^2f(x)}{dx^2} =  - N_f\frac{(f(x))^{3/2}}{\sqrt{x}}.\label{ddf}
\end{eqnarray}
Using Eq.(\ref{ddf}), the normalization integral may be integrated to
\begin{equation}
\int_0^{x_{max}} dx \sqrt{x}(f(x))^{3/2}=-\frac{1}{N_f}\left( x\frac{df}{dx}\right)_{x_{max}},
\end{equation}
which gives the endpoint differential condition
\begin{equation}
\left(\frac{df}{dx}\right)_{x_{max}}=-\frac{1}{x_{max}}=-\frac{\lambda R}{\frac{4}{3}g^2},
\label{dercond}
\end{equation}
where $\lambda$ is the Lagrange multiplier. Notice that the scaling substitutions,
\begin{equation}
f \longrightarrow f, x \longrightarrow \frac{x}{N_f^{2/3}}, \lambda \longrightarrow N_f^{2/3}\lambda\label{scaling}
\end{equation}
completely converts Eqs.(\ref{norm1})-(\ref{dercond}) into the equivalent single flavor system.

\begin{figure}
\begin{center}
\leavevmode
\includegraphics*[scale=.9]{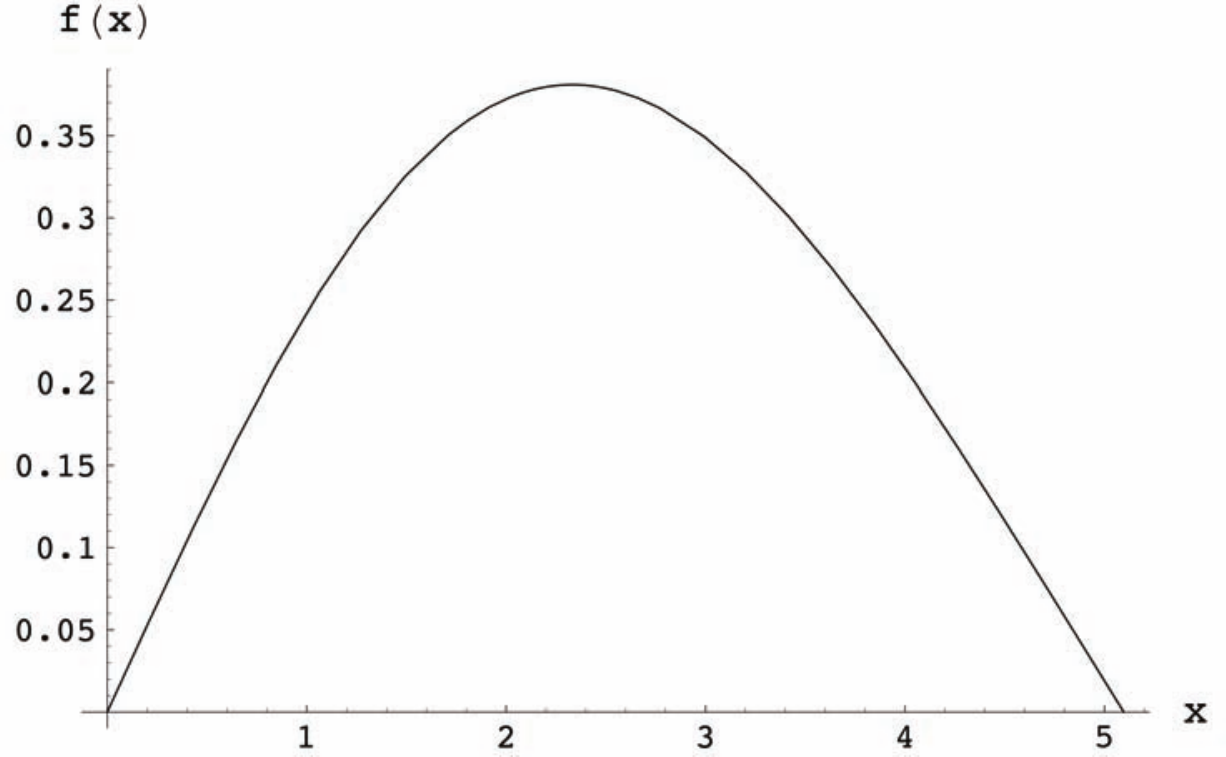}
\caption{The dimensionless TF spatial function, $f(x)$, related to the particle density by Eq.(\ref{FinN}), for the nonrelativistic $N_f=1$ flavor model. The dimensionless distance, $x$, is related to the physical radial coordinate, $r$, by Eqs.(\ref{vtrans}) and (\ref{peq}).}\label{TFfunction1}
\end{center}
\end{figure}

\begin{figure}
\begin{center}
\leavevmode
\includegraphics*[scale=.9]{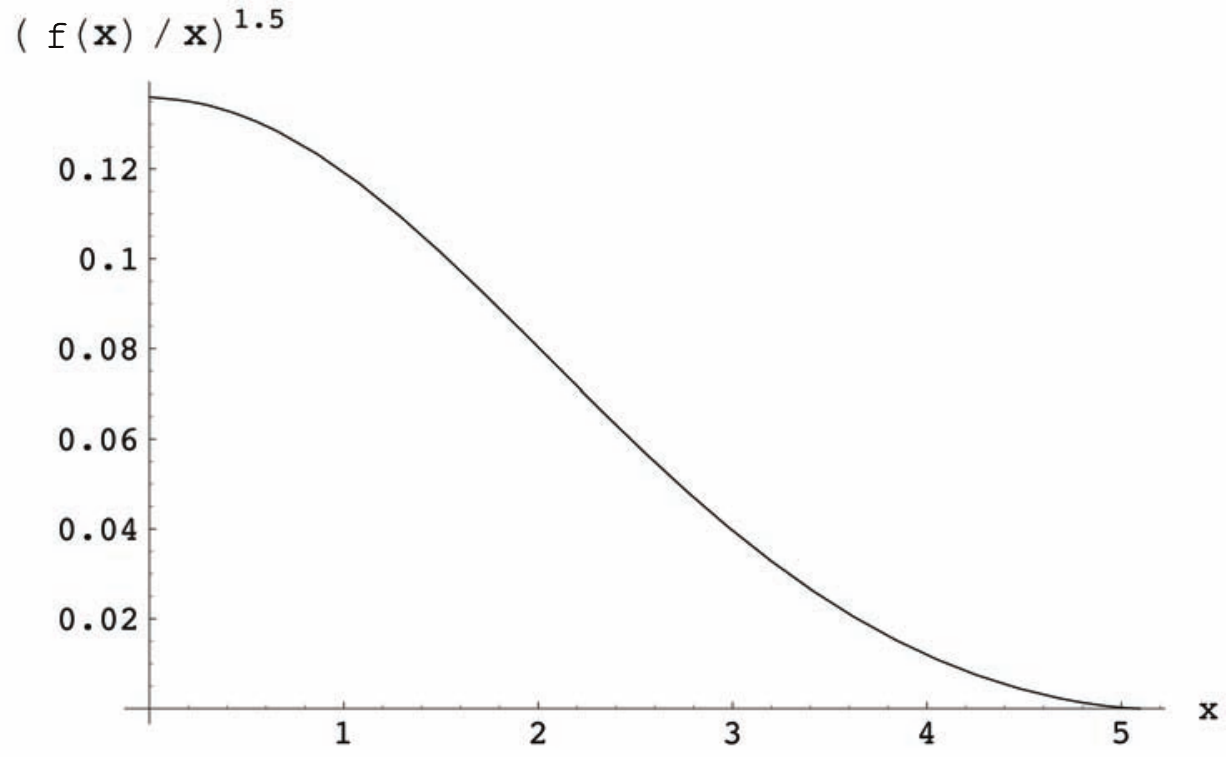}
\caption{The (unnormalized) particle density function, $(f(x)/x)^{3/2}$ (see Eq.(\ref{FinN})) for the nonrelativistic $N_f=1$ flavor model.}\label{normfig1}
\end{center}
\end{figure}

The equations of motion for this model may be solved numerically. The TF function obtained for $N_f=1$ is shown in Fig.~\ref{TFfunction1}. Since there is no central Coulombic source, the value of the function is zero at the origin, as can be understood from Eq.(\ref{f}). It extends out to $x_{max}=5.0965$. The particle density function, which is proportional to $(f(x)/x)^{3/2}$ (see Eq.(\ref{FinN})) is shown in Fig.~\ref{normfig1} also for the $N_f=1$ case. It has a smooth profile, even at the surface. Remember that the general $N_f$ case can be recovered from these results from the inverse of the (\ref{scaling}) substitutions.

I now wish to evaluate the kinetic and potential energies of this model. We will find that these can also be related to the endpoint derivative, $(df/dx)_{x_{max}}$. For the kinetic energy, $T$, we start with
\begin{equation}
T= N_f\int^{r_{max}}d^3r \frac{3(3\pi^2 \hbar^3 n(r))^{5/3}}{10\pi^2 \hbar^3 m}.
\end{equation}
In terms of the TF function, $f(x)$, one has
\begin{equation}
T= \frac{24N_f}{5\pi}\left(\frac{3\pi A}{4}\right)^{1/3}\frac{\frac{4}{3}g^2\times\frac{4}{3}\alpha_s}{a}
\int_0^{x_{max}}\frac{(f(x))^{5/2}}{\sqrt{x}},
\end{equation}
or, using Eq.(\ref{ddf}),
\begin{equation}
T= \frac{24}{7\pi}\left(\frac{3\pi A}{4}\right)^{1/3}\frac{\frac{4}{3}g^2\times\frac{4}{3}\alpha_s}{a}
\left(  x\left( \frac{df}{dx} \right)^2\right)_{x_{max}}.
\end{equation}
Finally, using Eq.(\ref{dercond}) we have the simple form
\begin{equation}
T= \frac{24}{7\pi}\left(\frac{3\pi A}{4}\right)^{1/3}\frac{\frac{4}{3}g^2\times\frac{4}{3}\alpha_s}{ax_{max}}=\frac{9}{7}\left(\frac{\frac{4}{3}g^2A}{r_{max}}\right).
\end{equation}

The potential energy, $U$, is given by
\begin{equation}
U = -\frac{3\times\frac{4}{3}g^2N_f^2}{2A}\int\int d^3r\,d^3r'\frac{n(r)n(r')}{|\vec{r}-\vec{r}\,'|}.\label{potenergy}
\end{equation}
This is probably the appropriate time to explain the philosophy behind the choice of energy/particle density normalization in the model. The discrete form of the Coulomb interaction between $3A$ colored objects is
\begin{equation}
U^{discrete} = -\frac{\frac{4}{3}g^2}{2\times(3A-1)}\sum_{i\ne j} \frac{1}{|\vec{r}_i-\vec{r}_j|}.
\label{discrete}
\end{equation}
We see in Eq.(\ref{discrete}) the residual interaction coupling, discussed in Section II, times the Coulomb interaction of charges with normalization ${\vec q}\,^2=\frac{4}{3}g^2$. The number of terms is $3A(3A-1)$, given the double counting in the sum. We may write Eq.(\ref{discrete}) as
\begin{equation}
U^{discrete} = -\frac{\frac{4}{3}g^2\times 3A}{2}<U>,
\label{discrete2}
\end{equation}
where $<U>$ represents the average interaction potential energy term. On the other hand, if the single particle norm $\hat{n} = n\frac{N_f}{A}$ ($\int d^3r \,\hat{n}(r) =1$) is used in (\ref{potenergy}), we have
\begin{equation}
U = -\frac{\frac{4}{3}g^2\times 3A}{2}\int\int d^3r\,d^3r'\frac{\hat{n}(r)\hat{n}(r')}{|\vec{r}-\vec{r}\,'|}.
\end{equation}
The comparison yields
\begin{equation}
<U> \leftrightarrow \int\int \frac{d^3r\,d^3r'}{|\vec{r}-\vec{r}\,'|}\hat{n}(r)\hat{n}(r').\label{compare}
\end{equation}
This is a natural association, which I consider the appropriate way to to set the overall energy normalization in the model. Of course, (\ref{discrete}) does not contain self-energy contributions whereas (\ref{potenergy}) does, so the comparison can not be exact. 

\begin{figure}
\begin{center}
\leavevmode
\includegraphics*[scale=.8]{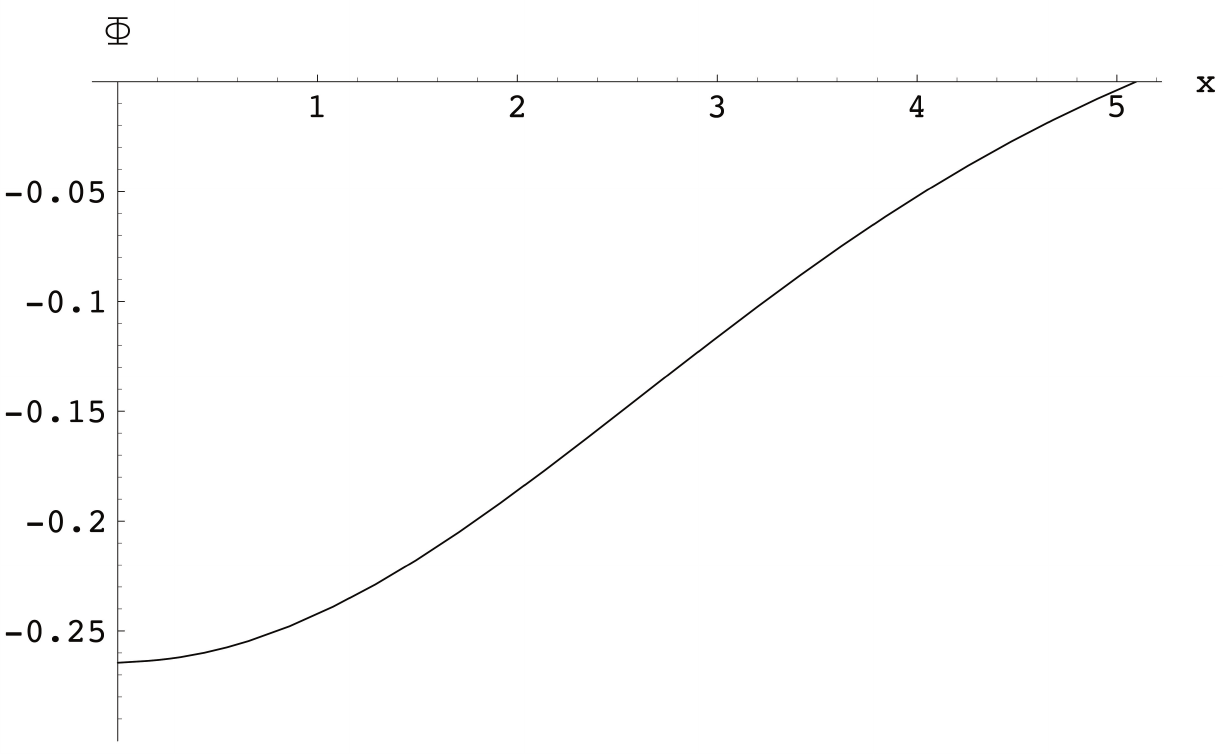}
\caption{The dimensionless potential, $\Phi\equiv -f(x)/x$, from Eq.(\ref{dimpot}), associated with the nonrelativistic $N_f=1$ flavor model.}\label{NRpot}
\end{center}
\end{figure}

I use Eq.(\ref{Eequation}) to define the potential, $V(r)$, in this model:
\begin{equation}
V(r) \equiv \lambda -\frac{\frac{4}{3}g^2N_f}{A}\int d^3r \frac{n(r)}{|{\vec r}-{\vec r}\,'|}.\label{potential}
\end{equation}
Evaluation gives
\begin{equation}
V(r)=-\frac{\frac{4}{3}g^2}{R}\frac{f(x)}{x},\label{dimpot}
\end{equation}
where I have used the definition of the Lagrange multiplier, Eq.(\ref{dercond}). Its role is to remove the discontinuity in the potential at the surface. The dimensionless form of this potential is displayed in Fig.~\ref{NRpot}. It shows an attractive collective potential with no central Coulombic spike.

Relating the potential energy to the TF function, $f(x)$, we have
\begin{eqnarray}
\nonumber U = -\frac{4N_f^2}{\pi}
\left(\frac{3\pi A}{4}\right)^{1/3}
\frac{\frac{4}{3}g^2\times\frac{4}{3}\alpha_s}{a}
\left[  \int_0^{x_{max}} \!\!dx \frac{(f(x))^{3/2}}{\sqrt{x}} 
\int_0^x dx' \sqrt{x'}(f(x'))^{3/2}\right. \\
\nonumber + \left.  \int_0^{x_{max}} \!\!dx  \sqrt{x}(f(x))^{3/2}
\int_x^{x_{max}} \!\!dx' \frac{(f(x'))^{3/2}}{\sqrt{x'}} \right].\\
\end{eqnarray}
The integrals yield
\begin{equation}
U=  -\frac{4}{\pi}
\left(\frac{3\pi A}{4}\right)^{1/3}
\frac{\frac{4}{3}g^2\times\frac{4}{3}\alpha_s}{a}
\left( \frac{12}{7}x\left( \frac{df}{dx} \right)^2\right)_{x_{max}}.
\end{equation}
Using Eq.(\ref{dercond}), we have
\begin{equation}
U=  -\frac{48}{7\pi}
\left(\frac{3\pi A}{4}\right)^{1/3}
\frac{\frac{4}{3}g^2\times\frac{4}{3}\alpha_s}{ax_{max}}=-\frac{18}{7}\left(\frac{\frac{4}{3}g^2A}{r_{max}}\right).
\end{equation}
Therefore
\begin{equation}
T=-\frac{1}{2}U,
\end{equation}
as we expect, and the energy of the system is just
\begin{equation}
E=-\frac{9}{7}\left(\frac{\frac{4}{3}g^2A}{r_{max}}\right).\label{f=0energy}
\end{equation}

The system energy is proportional to $A^{1/3}$, which means it is always energetically favorable to populate $n$ ($n=1, 2, 3\ldots$) $A=1$ states rather than a single $A=n$ state. Of course all these states would be occupied at nonzero temperature. The size of these loosely bound systems grows like $\sim A^{2/3}$, according to Eqs.(\ref{vtrans}) and (\ref{peq}). This is in contrast to the usual increase $\sim A^{1/3}$ for nuclear systems and is a result of the \lq\lq residual" nature of the coupling as discussed in Section II. Also note that although the number of degenerate flavors, $N_f$, does not appear explicitly in (\ref{f=0energy}), the scaling substitution, (\ref{scaling}), assures that the $N_f$ flavor energy, $E^{N_f}$, is related to the single flavor energy, $E^1$, by $E^{N_f}=N_f^{2/3}E^1$. Since all energies are negative, the $N_f$-degenerate system is more strongly bound than the single flavor system for the same value of $A$, as it should be.

Although interesting in themselves, it is clear that the states being formed do not constitute normal hadronic matter. As discussed in the Introduction, I interpret these states as loosely bound hadronic states in the zero temperature quark-gluon plasma phase. As a model of the confined phase, I next consider what happens to these solutions when a confining vacuum energy density term is added, as in the MIT quark model.

\section{Nonrelativistic $N_f$-Flavor Model with Volume Pressure}

I now introduce the volume energy,
\begin{equation}
E_{vol}=\frac{4\pi}{3}r_{max}^3B,
\label{vole}
\end{equation}
where $B$ is the MIT bag constant. The total energy is of course
\begin{equation}
E_{tot} = E_{vol}+T+U.\label{totalE}
\end{equation}

\begin{figure}
\begin{center}
\leavevmode
\includegraphics*[scale=.9]{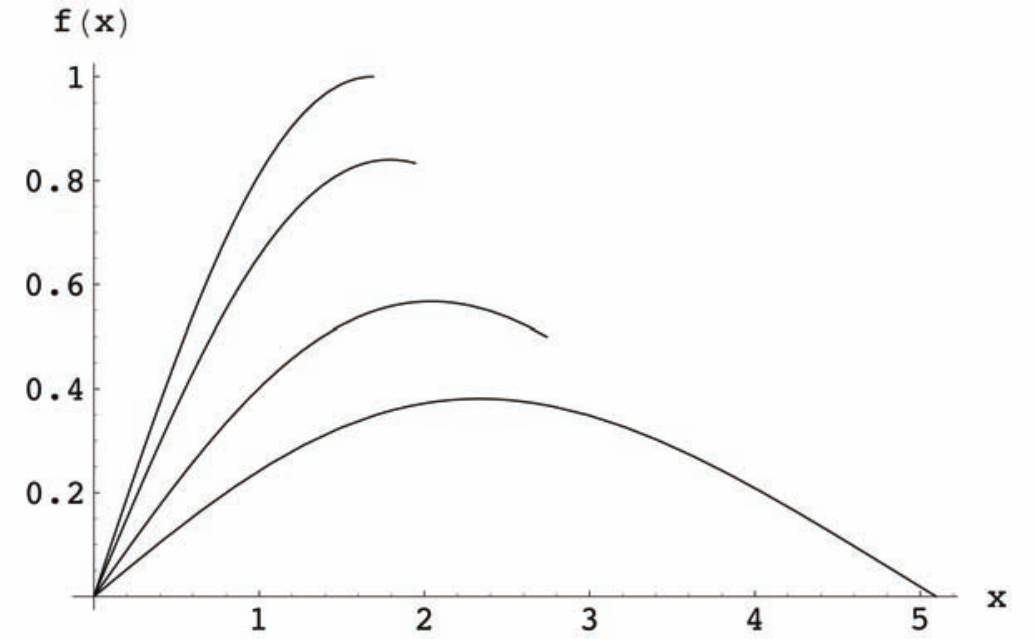}
\caption{The nonrelativistic TF functions, $f(x)$, as a function of $x$ for $N_f=1$. Reading from bottom to top, this shows the $f(x_{max})=0, \frac{1}{2},\frac{5}{6}$ and $1$ functions, corresponding to $x_{max}=5.097, 2.741, 1.945$ and $1.691$.}\label{TFmodel}
\end{center}
\end{figure}

\begin{figure}
\begin{center}
\leavevmode
\includegraphics*[scale=.9]{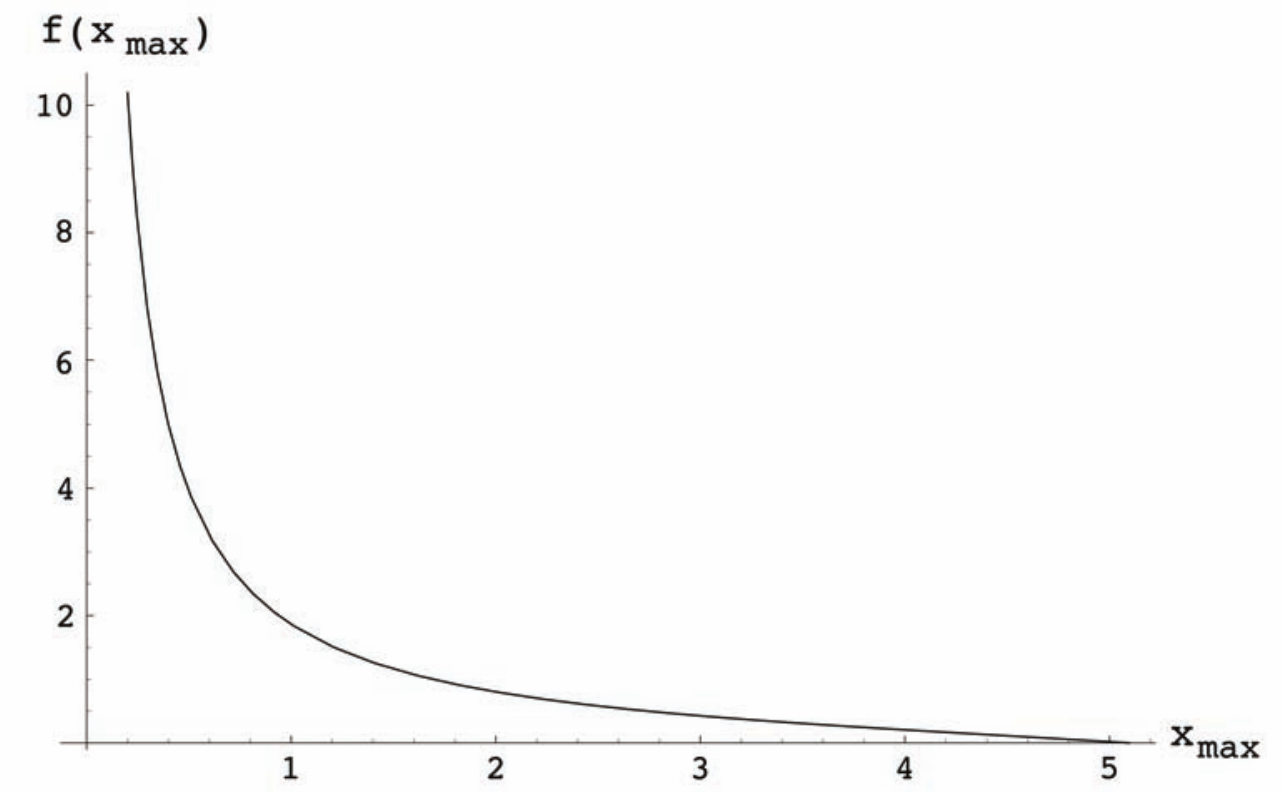}
\caption{ $f(x_{max})$ as a function of $x_{max}$ for the $N_f=1$ case.}\label{NRFvsx_max}
\end{center}
\end{figure}

We will again solve for the case of $N_f$ mass-degenerate quark flavors with equal numbers, $N^I$. One may anticipate that a discontinuity will be produced by the introduction of an external pressure, so that $f(x_{max}) \ne 0$. Given this, the normalization integral is now (use the differential equation (\ref{ddf}) in the integral and do an integration by parts)
\begin{equation}
\int_0^{x_{max}} dx \sqrt{x}(f(x))^{3/2}=\frac{1}{N_f}\left( f-x\frac{df}{dx}\right)_{x_{max}},
\end{equation}
where of course the integral is required to have the same value as in Eq.(\ref{norm1}). This leads to the modified boundary condition (also use Eq.(\ref{df}) to relate this to $\lambda$ at the boundary)
\begin{equation}
\left(\frac{df}{dx}\right)_{x_{max}}=\left(\frac{f-1}{x}\right)_{x_{max}}=-\frac{\lambda R}{\frac{4}{3}g^2}.\label{dercond1.5}
\end{equation}
These solutions can be characterized by the value of the TF function at the endpoint, $f(x_{max})$. These are displayed in Fig.~\ref{TFmodel} for the $N_f=1$ case. Note how the endpoint value of the TF function rises with decreased $x_{max}$. This is shown in Fig.~\ref{NRFvsx_max}.

Considering that there is a pressure-related discontinuity at the surface, the kinetic energy in this model is now given by
\begin{eqnarray}
T= \frac{24}{5\pi}\left(\frac{3\pi A}{4}\right)^{1/3}\frac{\frac{4}{3}g^2\times\frac{4}{3}\alpha_s}{a}  \left[ -\frac{5}{7}f\frac{df}{dx}  + \frac{5}{7}x\left( \frac{df}{dx} \right)^2
+ \frac{4N_f}{7}\sqrt{x}f^{5/2}\right]_{x_{max}}.
\end{eqnarray}

Using Eq.(\ref{dercond1.5})
\begin{equation}
T=\frac{24}{5\pi}\left(\frac{3\pi A}{4}\right)^{1/3}\frac{\frac{4}{3}g^2\times\frac{4}{3}\alpha_s}{a}
\left[    -\frac{5}{7}\frac{f-1}{x}  +\frac{4N_f}{7}\sqrt{x}f^{5/2}\right]_{x_{max}}.
\label{t}
\end{equation}
Likewise, for the potential energy, we have
\begin{eqnarray}
U=  -\frac{4}{\pi}
\left(\frac{3\pi A}{4}\right)^{1/3}
\frac{\frac{4}{3}g^2\times\frac{4}{3}\alpha_s}{a}
 \left[   -\frac{12}{7}f\frac{df}{dx} + \frac{12}{7}x\left( \frac{df}{dx} \right)^2
+ \frac{4N_f}{7}\sqrt{x}f^{5/2}\right]_{x_{max}}.
\end{eqnarray}
Again using (\ref{dercond1.5})
\begin{equation}
U=  -\frac{4}{\pi}
\left(\frac{3\pi A}{4}\right)^{1/3}
\frac{\frac{4}{3}g^2\times\frac{4}{3}\alpha_s}{a}
\left[   -\frac{12}{7}\frac{f-1}{x} +
\frac{4N_f}{7}\sqrt{x}f^{5/2}\right]_{x_{max}},
\end{equation}
and therefore
\begin{equation}
E= T+U=\frac{4}{\pi}
\left(\frac{3\pi A}{4}\right)^{1/3}
\frac{\frac{4}{3}g^2\times\frac{4}{3}\alpha_s}{a}
\left[ \frac{6}{7}\frac{f-1}{x}  +
\frac{4N_f}{35}\sqrt{x}f^{5/2}\right]_{x_{max}}.
\label{energy}
\end{equation}

\begin{figure}
\begin{center}
\leavevmode
\includegraphics*[scale=.9]{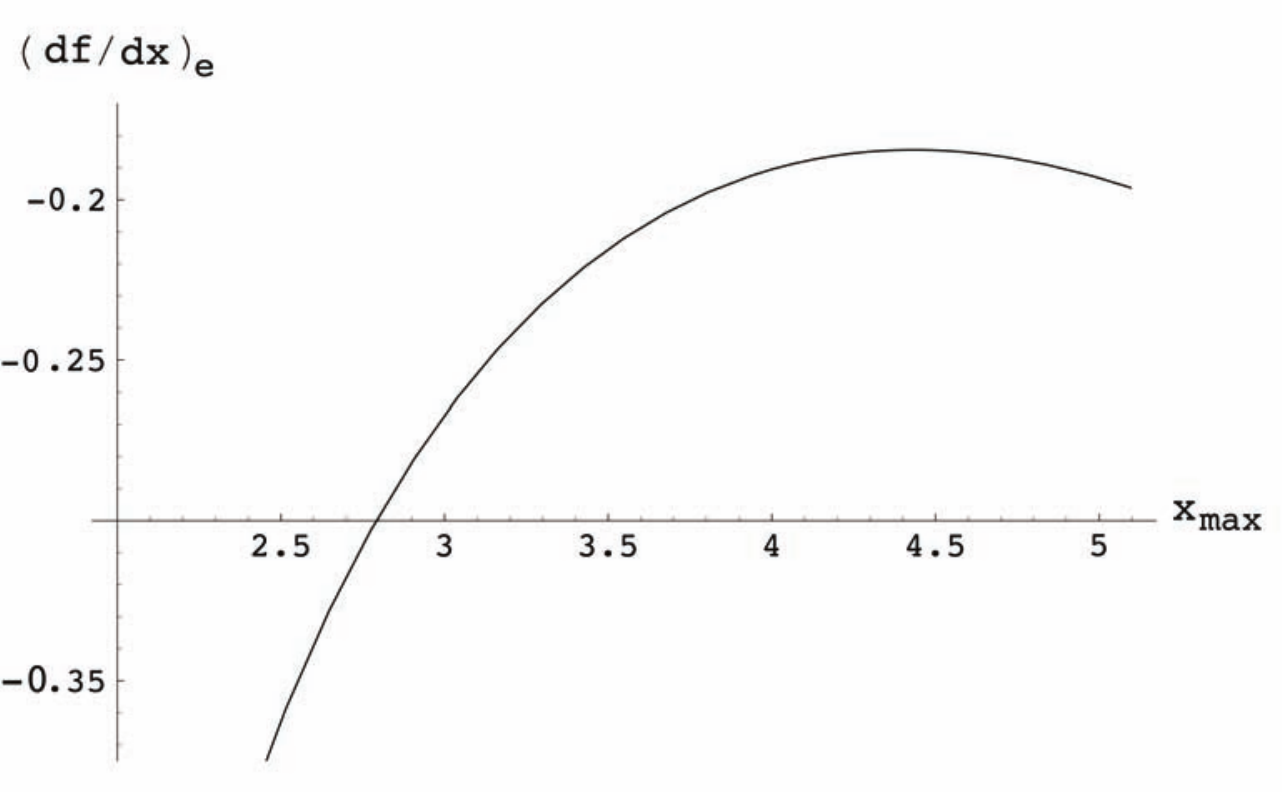}
\caption{The endpoint derivative of the function from Fig.~\ref{NRFvsx_max} in the region $2<x_{max}<x_f$. Note the maximum in the function near $x_{max}\simeq 4.5$.}\label{derivative}
\end{center}
\end{figure}

I wish to minimize the energy with respect to $r_{max}$, which, since $A$ is held constant, is the same as minimizing with respect to the dimensionless radius, $x_{max}$. In taking the derivative of the total energy, one has to be careful about the meaning of derivatives. One is taking the derivatives of $f(x_{max})$ with the normalization integral, Eq.(\ref{norm1}), being held constant. These special types of derivatives will be called \lq\lq endpoint" derivatives to distinguish them from the usual derivatives applied internally to the TF functions. They will be designated with a subscript, \lq\lq e". Using this notation, the pressure matching condition is
\begin{equation}
\left(\frac{\partial E_{tot}}{\partial x_{max}}\right)_e=0.
\end{equation}
The numerical relationship between $f(x_{max})$ and $x_{max}$ is given graphically in Fig.~\ref{NRFvsx_max} ($N_f=1$ case). Several limits of this figure may be easily understood. At the final point, $x_{max}=x_f$ ($x_f= 5.0965$ for $f(x_{f})=0$), one may show that $(df(x_{max})/dx_{max})_e|_{x_{max}=x_f}=-1/x_f$. The value of the endpoint derivative of the function in Fig.~\ref{NRFvsx_max}  is displayed in Fig.~\ref{derivative} for $2<x_{max}<x_f$. Note from this figure that there is a small maximum near the final point, $x_f$, leading to a small initial decrease in the slope. It is otherwise monotonically decreasing. In the other limit, $x_{max}\longrightarrow 0$, which we will see is obtained as $A\longrightarrow \infty$, one may show that
\begin{equation}
f(x_{max})\longrightarrow \left(\frac{3}{N_f}\right)^{2/3}\frac{1}{x_{max}}.\label{flimit}
\end{equation}

\begin{figure}
\begin{center}
\leavevmode
\includegraphics*[scale=.8]{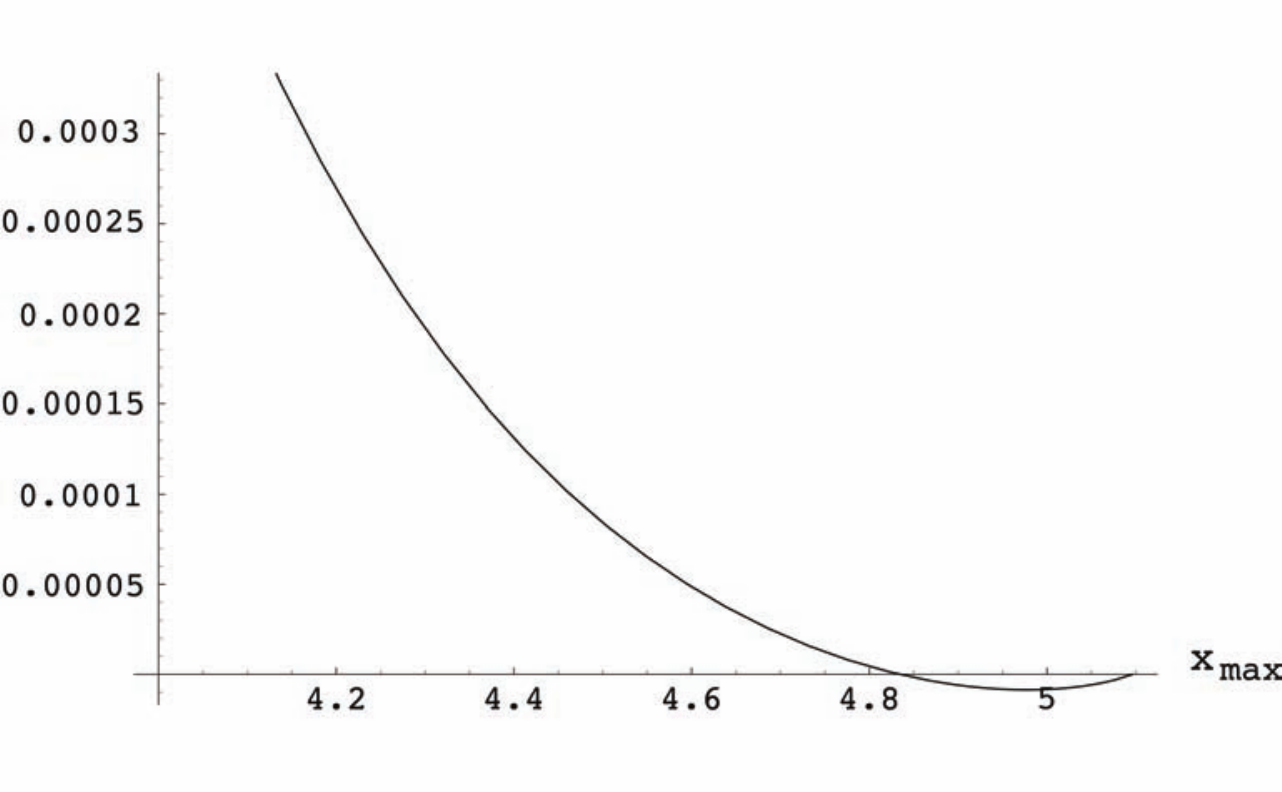}
\caption{ The \lq\lq matching function" of Eq.(\ref{xcon}) a function of $x_{max}$ for $N_f=1$ in the region $4<x_{max}<x_f$.}\label{matchingfig}
\end{center}
\end{figure}

Taking the derivatives of the various terms and setting them equal to zero to balance the pressure gives
\begin{eqnarray}
\nonumber \frac{\pi^2}{8}\left(\frac{Ba^4/\hbar c}{\alpha_s^5}\right) \left( \frac{3\pi A}{4} \right)^{5/3}=
\left[ \frac{1}{x^{2}} \left( \frac{6}{7}\frac{f-1}{x^2}-\frac{6}{7x}\left( \frac{df}{dx}\right)_e -\frac{2N_f}{35} \frac{f^{3/2}}{\sqrt{x}} \left( f+5x\left( \frac{df}{dx}\right)_e  \right)  \right) \right]_{x_{max}} ,\\  
\label{xcon}
\end{eqnarray}
where again we encounter endpoint derivatives. I will call this the \lq\lq matching equation" and the right-hand side the \lq\lq matching function". It is displayed in Fig.~\ref{matchingfig} for $4<x_{max}<x_f$. Note that the function is initially negative in the vicinity of $x_{max}=x_f$. This means there are no stable solutions in this region. This instability is a consequence of the small initial decrease in slope seen in Fig.~\ref{derivative}. The negative excluded region decreases as $N_f$ increases. It excludes the approximate regions $4.83<x_{max}<x_f$ for $N_f=1$, $5.03<x_{max}<x_f$ for $N_f=2$ and $5.07<x_{max}<x_f$ for $N_f=3$. Although it continues to shrink, the excluded region seems to persist for any value of $N_f$. Truly heavy quark systems, such as for charmed or bottom, would have small \lq\lq $a$" values that would force them to have $x_{max}$ values near the edge of this region. Finally, notice that the left-hand side of Eq.(\ref{xcon}) is obviously a very sensitive function of both $a$ and $\alpha_s$. 

Using Eq.(\ref{xcon}) in (\ref{vole}) allows one to formally eliminate the bag constant $B$, and gives the total energy as
\begin{eqnarray}
\nonumber E_{tot}&=&\frac{4}{\pi}\left(\frac{3\pi A}{4} \right)^{1/3}\left( \frac{\frac{4}{3}g^2\times \frac{4}{3}\alpha_s}{a} \right) \\
&&\times \left(  \frac{8}{7} \frac{f-1}{x} -\frac{2}{7}\left( \frac{df}{dx} \right)_e +  \frac{2N_f}{21}\sqrt{x}f^{3/2}\left(f-x\left(\frac{df}{dx}\right)_e \right)  \right)_{x_{max}}\label{etot}
\end{eqnarray}
Note that $E_{tot}=0$ is not the reference energy for a system being bound, since there are no free particle states. One can only compare the energies of one set of bound states with another in the confined phase.

The energies as a function of $A$ from Eq.(\ref{etot}) are complicated and they have not yet been explored. However, in the limit of $A\longrightarrow \infty$, one can show from Eqs.(\ref{flimit}) and (\ref{xcon}) that the system shrinks, $x_{max}\sim A^{-1/3}$, leading to $E_{tot} \sim A$. However, this limit is unphysical because the system is becoming relativistic as it is being confined to a smaller volume. This leads us to our next set of considerations.

\section{Relativistic Models}

Can the present model be made relativistic? Although the expression for the relativistic energy density, $({\cal E}_R)^I_i$, is more complicated, the equation replacing Eq.(\ref{Eequation}) above is clear:
\begin{eqnarray}
\nonumber \sqrt{(p_F^Ic)^2+(m_Ic^2)^2} -m_Ic^2&=&-\lambda^I_R +
\frac{3\times\frac{4}{3}g^2}{(3A-1)}\left(\frac{N^I-1}{N^I} \int^{r_{max}}\!\!d^3r 
\frac{n^I(r')}{|{\vec r}-{\vec r}\,'|} \right. \\
&+&\left. \sum_{J\ne I}\int^{r_{max}}\!\!d^3r' \frac{n^J(r')}{|{\vec r}-{\vec r}\,'|} \right),
\end{eqnarray}
with $p_F^I$ again given in terms of $n^I$ by Eq.(\ref{np_F}). This reduces to the previous nonrelativistic equation in the low density limit. Instead, I will concentrate on showing the existence and properties of the extreme relativistic or massless version of this equation. This is given by
\begin{equation}
p_F^Ic=-\lambda^I_R +
\frac{3\times\frac{4}{3}g^2}{(3A-1)}\left(\frac{N^I-1}{N^I} \int^{r_{max}}\!\!d^3r 
\frac{n^I(r')}{|{\vec r}-{\vec r}\,'|} 
+\sum_{J\ne I}\int^{r_{max}}\!\!d^3r' \frac{n^J(r')}{|{\vec r}-{\vec r}\,'|} \right).
\end{equation}
I will continue to assume $N_f$ degenerate flavors with equal numbers, $N^I$, in this section. Quantities such as $p^I_F$, $\lambda_R^I$, and $n^I(r)$ now lose their flavor superscript. I will introduce the notation $w(r)$ for the degenerate massless TF function. It is dimensionless and is related to the previous definition of $f(r)$, assuming $m\ne 0$, by
\begin{equation}
w(r)\equiv \frac{r}{\frac{4}{3}\alpha_s}(3\pi^2n(r))^{1/3}=\left(\frac{2mc^2f(r)r}{\frac{4}{3}g^2}\right)^{1/2}.\label{wdef}
\end{equation}
Eq.(\ref{singlenorm}) now reads
\begin{equation}
\int^{r_{max}} \frac{dr}{r}\,(w(r))^3 = \frac{1}{\beta},
\label{newnorm}
\end{equation}
where 
\begin{equation}
\beta\equiv \frac{4(\frac{4}{3}\alpha_s)^3N_f}{3\pi A}.
\end{equation}

We follow a path similar to the one for the nonrelativistic model. The TF integral equation in this case reads
\begin{equation}
w(r) =   -\frac{\lambda_R r}{\frac{4}{3}g^2} + \beta\left[ \int_0^r \frac{dr'}{r'}\left(w(r')\right)^3+  r \int_r^{r_{max}}\!\!\frac{dr'}{r'\,^2} \left(w(r')\right)^3\right] \label{w}.
\end{equation}
The first derivative is
\begin{equation}
\frac{dw}{dr} =   -\frac{\lambda_R}{\frac{4}{3}g^2} + \beta\int_r^{r_{max}}\!\!\frac{dr'}{r'\,^2} \left(w(r')\right)^3,
\end{equation}
while the second is
\begin{equation}
\frac{d^2w}{dr^2} =   -\beta\frac{w^3}{r^2}\label{diffeq2}.
\end{equation}
\begin{figure}
\begin{center}
\leavevmode
\includegraphics*[scale=.9]{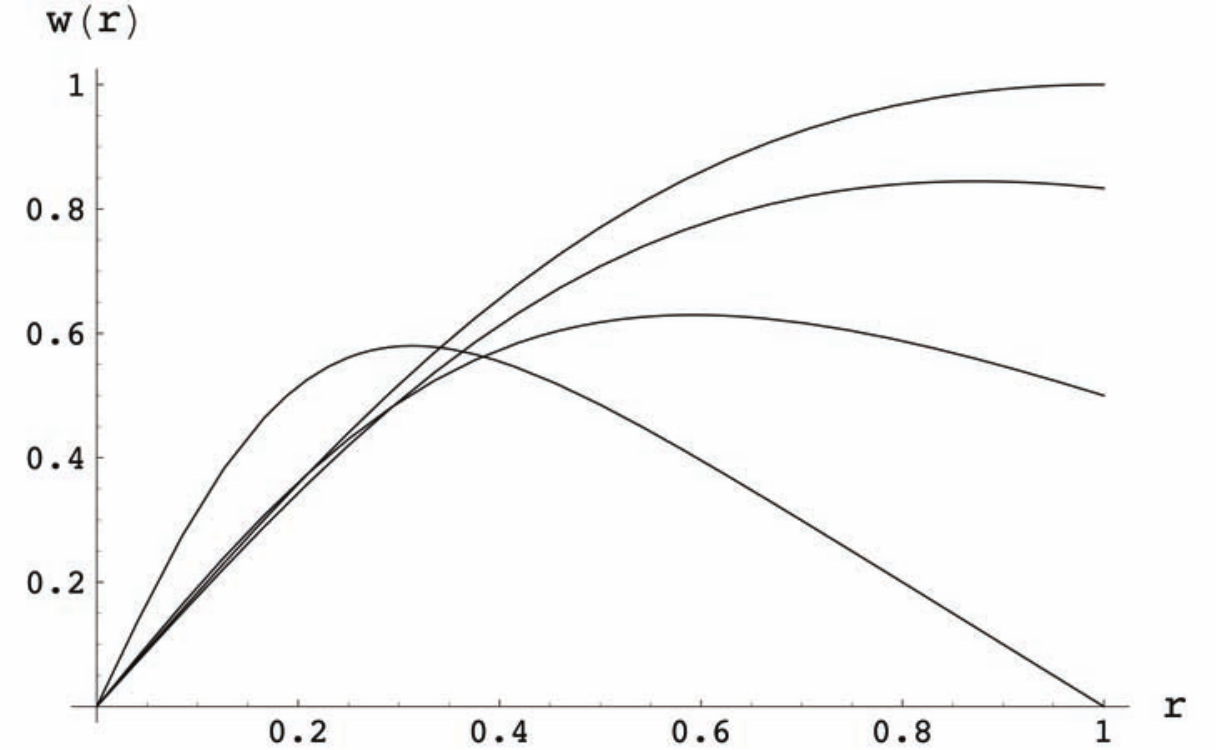}
\caption{The relativistic TF wavefunctions as a function of $x$. Reading from bottom to top, this shows the $w(r_{max})=0, \frac{1}{2},\frac{5}{6}$ and $1$ functions with $r_{max}=1$.}\label{Rfunctions}
\end{center}
\end{figure}
Eq.(\ref{newnorm}) nows gives
\begin{equation}
\left(\frac{dw}{dr}\right)_{r_{max}}=\left(\frac{w-1}{r}\right)_{r_{max}}=-\frac{\lambda_R}{\frac{4}{3}g^2},
\label{dercond2}
\end{equation}
similar to Eq.(\ref{dercond}). A version of Eq.(\ref{diffeq2}) appears in Ref.\cite{dixon} in the context of a study of the quark-gluon plasma with $SU(2)$ or $SU(3)$ color. In this paper the authors subject the partition function to a TF approximation and confine the system of massless quarks with an external, spherical box in order to study its thermodynamics. Instead of the single function $w(r)$, this development has two potentials, called $f(r)$ and $a(r)$, for $SU(2)$. If the gauge particles had abelian $U(1)$ symmetry, as is assumed here, the $a(r)$ function would not arise. The $f(r)$ equation would then go over to a scaled version of Eq.(\ref{diffeq2}) in the zero temperature, zero density (chemical potential) limit. However, the boundary conditions used for the field functions are quite different from those used here. Presumably, the equations presented in Ref.\cite{dixon} will not contain finite quark matter solutions since $\beta$ is replaced with a negative constant, producing repulsion, and $f(r)$ is seen to diverge for large distances. Perhaps requiring the plasma to be an overall color singlet, as is done here, would change the sign of this coefficient.

Getting back to Eq.(\ref{diffeq2}), notice that it is invariant under the substitution $r\longrightarrow Sr$, where $S$ is a scale factor. Thus Eq.(\ref{w}) formally has a solution, but no scale!  Unlike the nonrelativistic model which has the Compton length as an intrinsic scale, using $w(r_{max})=0$ as a boundary condition is meaningless here. This will be overcome by introducing the bag constant, which, like the nonrelativistic case, will result in a discontinuity in the TF function at the boundary. Formal solutions for the TF function $w(r)$ are shown in Fig.~\ref{Rfunctions}, where I have set $r_{max}=1$ in arbitrary units.

The ultra-relativistic kinetic energy is given by
\begin{equation}
T= N_f\int^{r_{max}}\!\!d^3r \frac{3c(3\pi^2 \hbar^3 n(r))^{4/3}}{4\pi^2 \hbar^3},
\end{equation}
or, when related to $w(r)$, we have
\begin{equation}
T= \frac{3c\hbar N_f}{\pi}(\frac{4}{3}\alpha_s)^4\int_0^{r_{max}}\!\!\frac{dr}{r^2} (w(r))^4.
\end{equation}
Evaluation, using Eq.(\ref{diffeq2}), relates this integral to endpoint values and derivatives of the $w$ function:
\begin{equation}
T= \frac{9}{2}(\frac{4}{3}g^2) A\left( -w\frac{dw}{dr}+r\left(\frac{dw}{dr}\right)^2 +\frac{\beta}{2} \frac{w^4}{r}\right)_{r_{max}}.
\end{equation}
Using Eq.(\ref{dercond2}) then gives
\begin{equation}
T= \frac{9}{2}(\frac{4}{3}g^2) A\left(\frac{1-w+\beta w^4/2}{r} \right)_{r_{max}}.
\end{equation}

\begin{figure}
\begin{center}
\leavevmode
\includegraphics*[scale=.9]{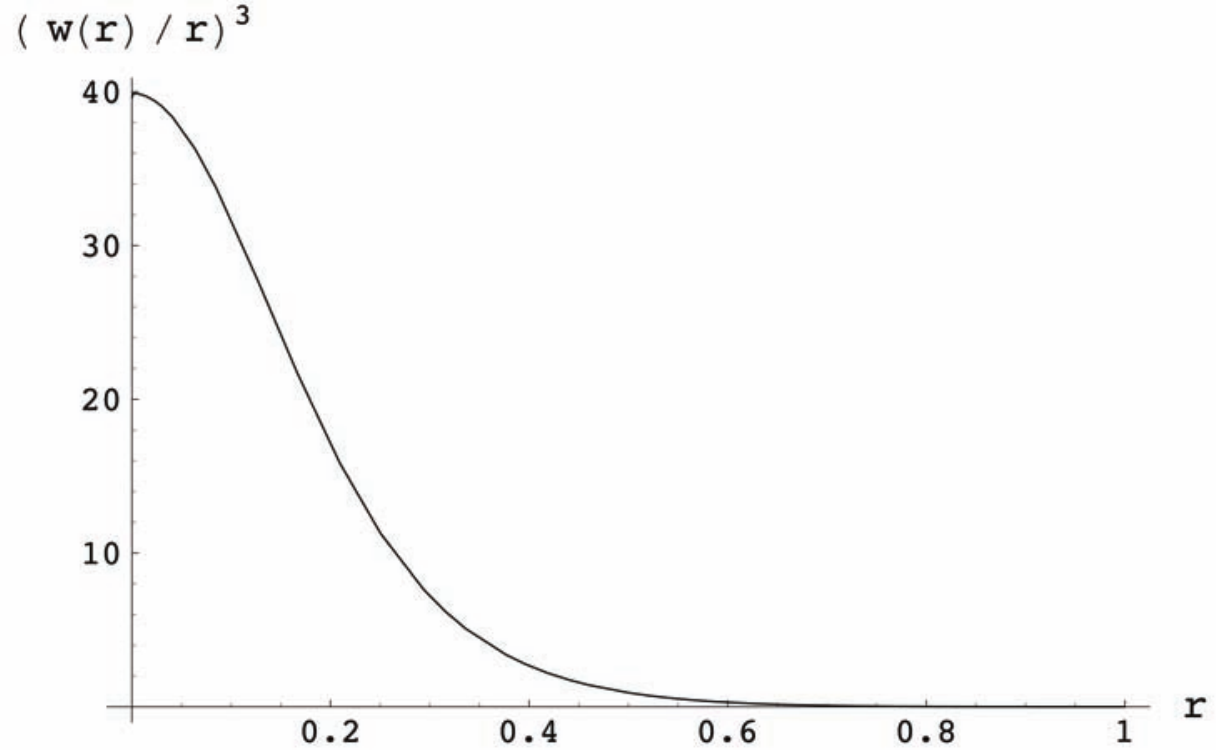}
\caption{The (unnormalized) relativistic TF density, proportional to $(w(r)/r)^3$ from Eq.(\ref{wdef}), for the scaleless $w(r_{max})=0$ model with $r_{max}=1$.}\label{Rdensity}
\vspace{1cm}
\includegraphics*[scale=.9]{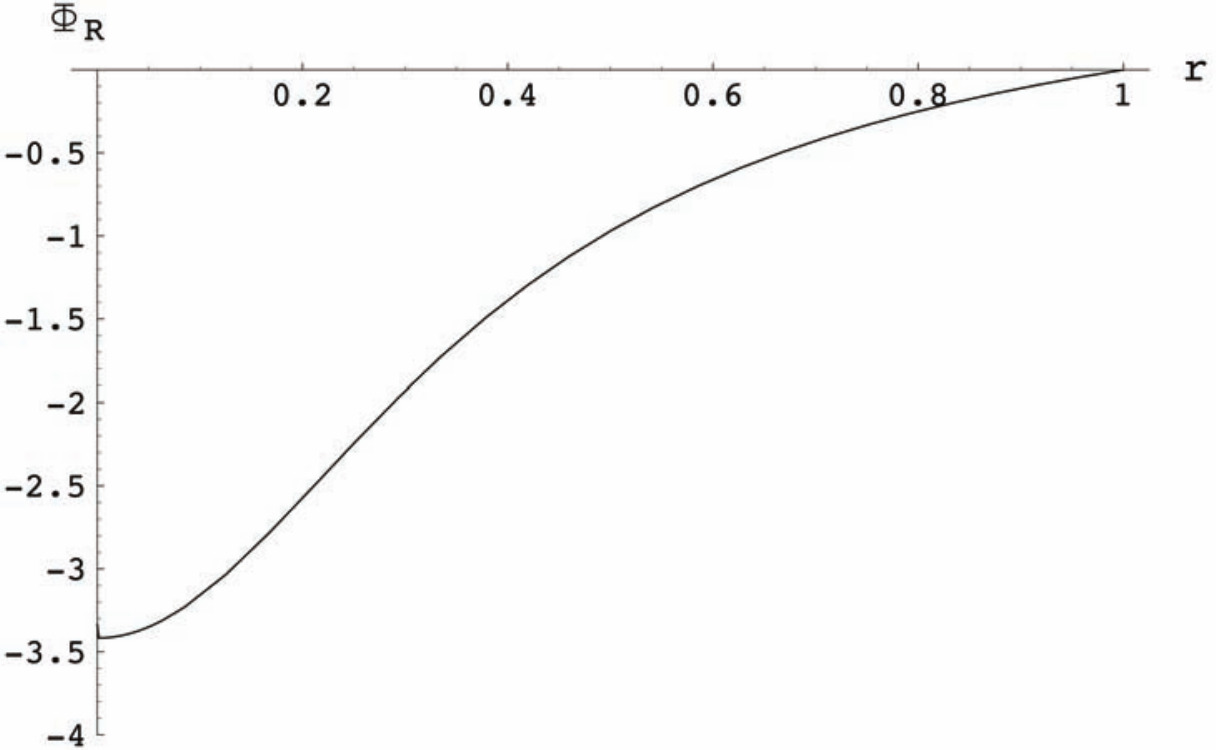}
\caption{The relativistic TF potential, $\Phi_R \equiv -w(r)/r$, from Eq.(\ref{VR}), for the scaleless $w(r_{max})=0$ model with $r_{max}=1$.}\label{Rpotential}
\end{center}
\end{figure}

It is interesting to examine the relativistic potential, $V_R(r)$. Similar to Eq.(\ref{potential}) above, one has
\begin{equation}
V_R(r)\equiv \lambda_R -\frac{\frac{4}{3}g^2N_f}{A}\int d^3r' \frac{n(r')}{|\vec{r}-\vec{r}\,'|}.
\end{equation}
This results in
\begin{equation}
V_R(r) = -\frac{4}{3}g^2 \frac{w(r)}{r}.\label{VR}
\end{equation}
The relativistic Lagrange multiplier, $\lambda_R$, plays the same role here as in the nonrelativistic model. The density function and the potential are displayed for the $w(r_{max})=0$ version of the model in Figs.~\ref{Rdensity} and \ref{Rpotential}, setting $r_{max}=1$ in arbitrary units. One again sees smooth density and potential profiles with no central Coulombic charge. Of course, the $w(r_{max})\ne 0$ models inevitably involve a discontinuity in these quantities at the surface.

The potential energy is (formally the same as Eq.(\ref{potenergy}))
\begin{equation}
U = -\frac{3\times\frac{4}{3}g^2N_f^2}{2A}\int\int d^3r\,d^3r'\frac{n(r)n(r')}{|\vec{r}-\vec{r}\,'|}.
\end{equation}
When related to the TF $w$ function, this becomes
\begin{eqnarray}
\nonumber U = -\frac{8}{3\pi^2}  \frac{c\hbar N_f^2}{A}(\frac{4}{3}\alpha_s)^7
 \left[   \int_0^{r_{max}} \!\!dr \frac{(w(r))^3}{r^2}
\int_0^r dr' \frac{(w(r'))^{3}}{r'}\right. \\
+ \left.  \int_0^{r_{max}} \!\!dr  \frac{(w(r))^3}{r}
\int_r^{r_{max}} \!\!dr' \frac{(w(r'))^3}{r'^2} \right].
\end{eqnarray}
Using techniques similar to the nonrelativistic evaluations, one obtains
\begin{eqnarray}
U = -\frac{9}{2}(\frac{4}{3}g^2) A
\left(-w\frac{dw}{dr}+r\left(\frac{dw}{dr}\right)^2 +\frac{\beta}{3} \frac{w^4}{r}\right)_{r_{max}}.
\end{eqnarray}
Finally, using Eq.(\ref{dercond2}), this is equivalent to
\begin{equation}
U= -\frac{9}{2}(\frac{4}{3}g^2) A\left(\frac{1-w+\beta w^4/3}{r} \right)_{r_{max}}.
\end{equation}
The total energy of the unconfined system is given by
\begin{equation}
E=T+U=\frac{c\hbar N_f}{\pi}(\frac{4}{3}\alpha_s)^4 \left(\frac{w^4}{r}\right)_{r_{max}}.\label{renergy1}
\end{equation}

\begin{figure}
\begin{center}
\leavevmode
\includegraphics*[scale=.9]{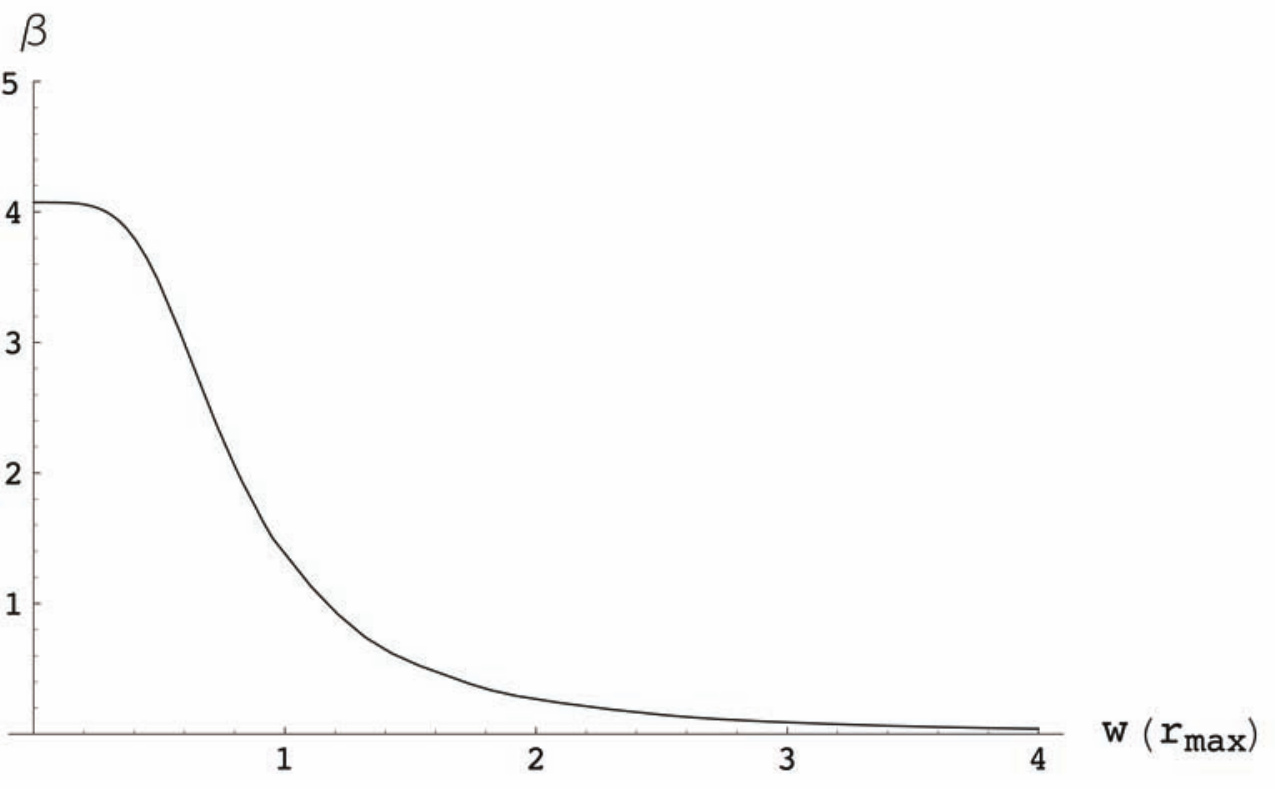}
\caption{$\beta$ as a function of $w(r_{max})$.}\label{Rbetavsw_max}
\end{center}
\end{figure}

The situation here is intrinsically different from the nonrelativistic model. The solution to Eq.(\ref{diffeq2}) with endpoint boundary conditions given by Eq.(\ref{dercond2}), for arbitrary $r_{max}$, shows that there is a one-to-one relationship between $\beta$ and $w(r_{max})$. This connection occurs of course because one is keeping the normalization integral, Eq.(\ref{newnorm}), constant as the surface discontinuity is varied. One finds that $\beta_f\equiv 4.073$ corresponds to $w(r_{max})=0$. This is illustrated in Fig.~\ref{Rbetavsw_max} where $\beta$ is plotted as a function of $w(r_{max})$. Notice that the function in this figure has zero slope at $w(r_{max})=0$. In the other limit, $w_{max}\longrightarrow 0$, which is obtained as $A\longrightarrow \infty$, one can show that
\begin{equation}
\beta\longrightarrow \frac{3}{w_{max}^3}.\label{limit2}
\end{equation}

For the confined system we define the total energy as in Eqs.(\ref{vole}) and (\ref{totalE}). I again seek the minimum energy solution by varying the radius, $r_{max}$ while keeping $A$, and therefore $w(r_{max})$,  constant. This gives
\begin{equation}
\left(\frac{\partial E_{tot}}{\partial r_{max}}\right)_e= 4\pi B r_{max}^2-\frac{c\hbar N_f}{\pi}(\frac{4}{3}\alpha_s)^4\left( \frac{w^4}{r^2 }\right)_{r_{max}}.
\end{equation}
By requiring this to be stationary one obtains a linear relation between $r_{max}$ and $w(r_{max})$:
\begin{equation}
r_{max}=\frac{4}{3}\alpha_s\left(\frac{N_f}{4\pi^2} \frac{c\hbar}{B}  \right)^{1/4}w(r_{max}).\label{radius}
\end{equation}
Notice that the $A$ value does not appear explicitly here, but is implicitly involved because of the connection between $w(r_{max})$ and $\beta$. The radius formula is no longer a matching condition to be solved, as in the nonrelativistic case, but simply a pressure matching result for the radius. After eliminating the bag constant, the total energy can be written as
\begin{equation}
E_{tot}= \frac{4c\hbar N_f}{3\pi}(\frac{4}{3}\alpha_s)^4\left(  \frac{w^4}{r}\right)_{r_{max}}.\label{relenergy}
\end{equation}
Energies here are intrinsically positive. The relationship between $\beta$ and $w(r_{max})$ implies a nontrivial $A$ dependence in Eq.(\ref{relenergy}) from the relationship between $w(r_{max})$ and $\beta$. Again, I have not yet explored the $A$ dependence of the energy in general. However, in the the $A\longrightarrow \infty$ limit, Eq.(\ref{limit2}) gives $E_{tot}\sim \beta^{-1}$, which shows that the total energy is proportional to $A$ and inversely proportional to $N_f$.

\section{Color-Flavor Locking Considerations}

One of the interesting possible applications of the present model would be to study so-called quark strangelets\cite{witten,farhi}, hypothetical finite lumps of hadronic matter made stable by an optimal mixture of up, down and strange quarks. Much work on such systems has already been done, especially in bag models\cite{BandJ}. The question of stability is critical, and comparison of the TF quark model with previous results should provide a consistency check.

In a very interesting set of papers, the authors in Ref.\cite{cflock} (see also the review articles\cite{cfreview}) have given evidence of a scenario in which the ground state of cold, low density quark matter is described by a mechanism called color-flavor locking (CFL). It is conceivable that for a large number of quarks this could produce finite quark matter\cite{madsen3}. CFL is due to an attractive instanton or single gluon exchange interaction between two quarks with different flavors and colors, and results in Cooper pairing and a diquark condensate. It also leads to an equal density of up, down and strange quarks, at least in the zero quark mass limit. Moreover, the spontaneous symmetry breaking involved leads to massive gluons via the Higgs mechanism. I will examine some aspects of the CFL modifications on the massless quark solutions of the last section, although these changes could also be made in the context of the nonrelativistic model of Sections IV and V.

Let us define $\mu\equiv m_gc/\hbar$, where $m_g$ is the gluon mass. I assume $N_f=3$ degenerate flavors and introduce a Cooper pairing gap, $\Delta$, in the momentum spectrum of the massless quarks. The formation of Cooper pairs leads to an approximate energy density\cite{energy},
\begin{equation}
{\cal E}_{pair}= -\frac{3\Delta^2 p_F^2}{c\hbar^3\pi^2},
\end{equation}
where $\Delta$ is the energy gap. In the TF model, such a term becomes a function of the color-flavor density function, $n(r)$, using Eq.(\ref{n}), and can be incorporated in the energy functional.

The modifications to Eq.(\ref{w}) of the last section are now straightforward. We find
\begin{eqnarray}
\nonumber w(r)-\frac{2\Delta^2}{3(\frac{4}{3}g^2)^2}\frac{r^2}{w(r)}  = -\frac{\lambda^* r}{\frac{4}{3}g^2} + \frac{\beta}{\mu}\left[ \int_0^r \frac{dr'}{r'\,^2}\left(w(r')\right)^3e^{-\mu r}\sinh(\mu r')  \right. \\
\left. + r \int_r^{r_{max}}\!\!\frac{dr'}{r'\,^2} \left(w(r')\right)^3e^{-\mu r'}\sinh(\mu r)\right]. \label{wYuk1}
\end{eqnarray}
Of course, the $r$ dependent quantities may be taken outside the integrals in (\ref{wYuk1}). One does not expect a singularity on the left-hand side from the pairing term due to the linear vanishing of $w(r)$ at the origin, and solutions with $w(r_{max})=0$ are avoided when the bag constant is employed. As in the last section, I will bypass a general consideration of this equation in favor of a special case. We will concentrate on looking at the massive gluon modifications and set $\Delta=0$ in the following.

The first derivative of Eq.(\ref{wYuk1}) for $\Delta=0$ gives
\begin{eqnarray}
\nonumber \frac{dw}{dr} =   -\frac{\lambda^* }{\frac{4}{3}g^2} + \beta \left[ -\int_0^r \frac{dr'}{r'\,^2}\left(w(r')\right)^3e^{-\mu r}\sinh(\mu r')\right. \\
\left. +  r \int_r^{r_{max}}\!\!\frac{dr'}{r'\,^2} \left(w(r')\right)^3e^{-\mu r'}\cosh(\mu r)\right].\label{wYuk2}
\end{eqnarray}
\begin{figure}
\begin{center}
\leavevmode
\includegraphics*[scale=.9]{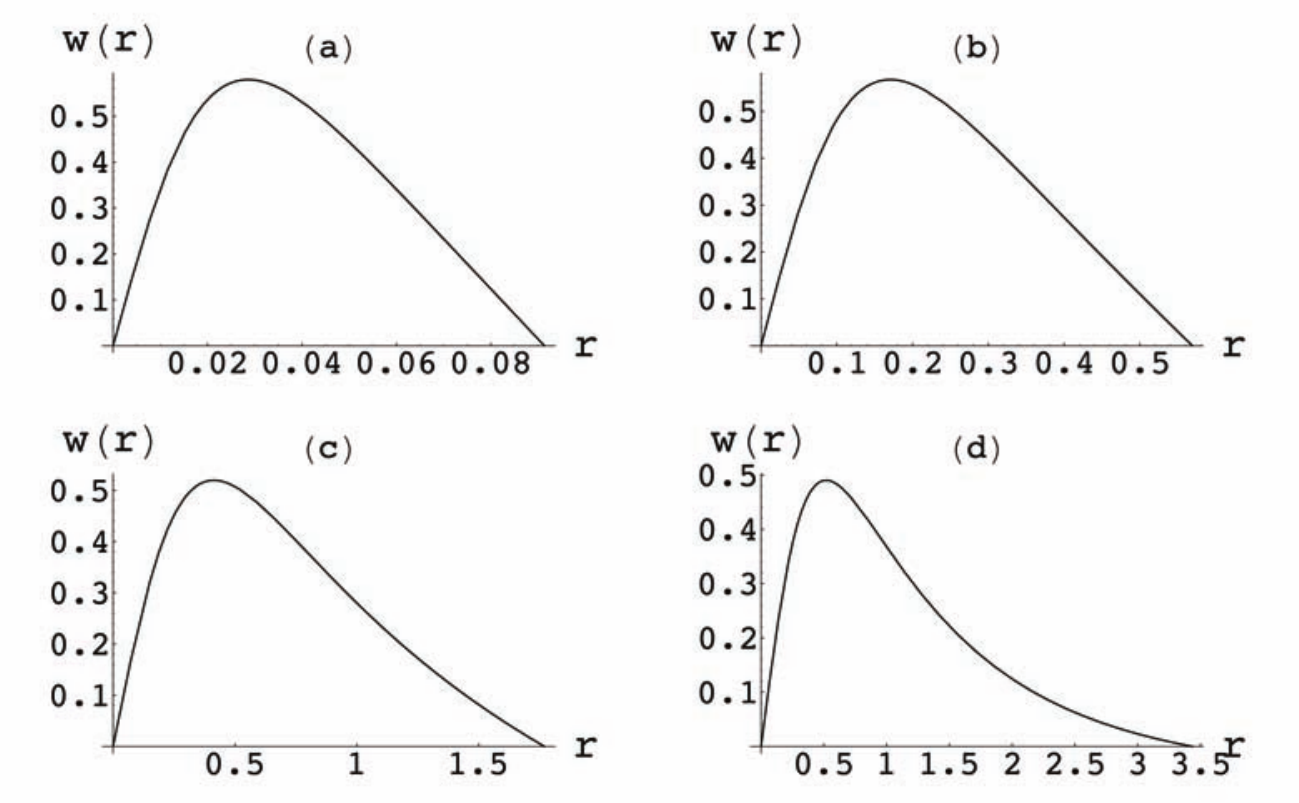}
\caption{TF massive gluon functions, $w(r)$, for (a)$\lambda^*=10$ (b)$\lambda^*=1$ (c)$\lambda^*=0.1$ (d)$\lambda^*=0.01$. $\lambda^*$ in units of $\hbar c\mu$ and distances measured in units of $\mu^{-1}$.}\label{Ytableau}
\end{center}
\end{figure}
\begin{figure}
\begin{center}
\leavevmode
\includegraphics*[scale=.8]{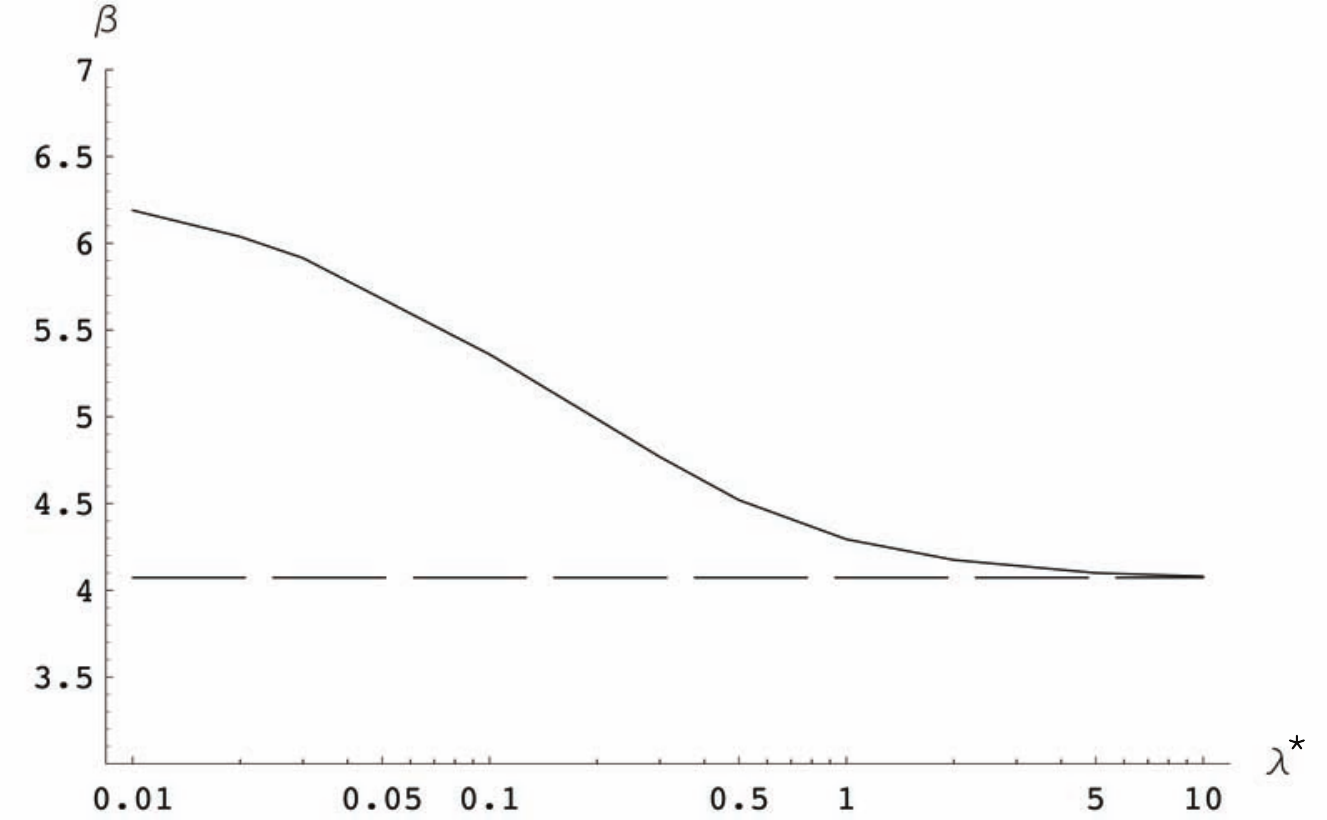}
\caption{$\beta$ as a function of $\lambda^*$, measured in units of $\hbar c\mu$, for the massive gluon model. The dashed horizonal line gives the asymptote $\beta_f=4.0733$.}\label{Ybetavslambda}
\end{center}
\end{figure}
\begin{figure}
\begin{center}
\leavevmode
\includegraphics*[scale=.8]{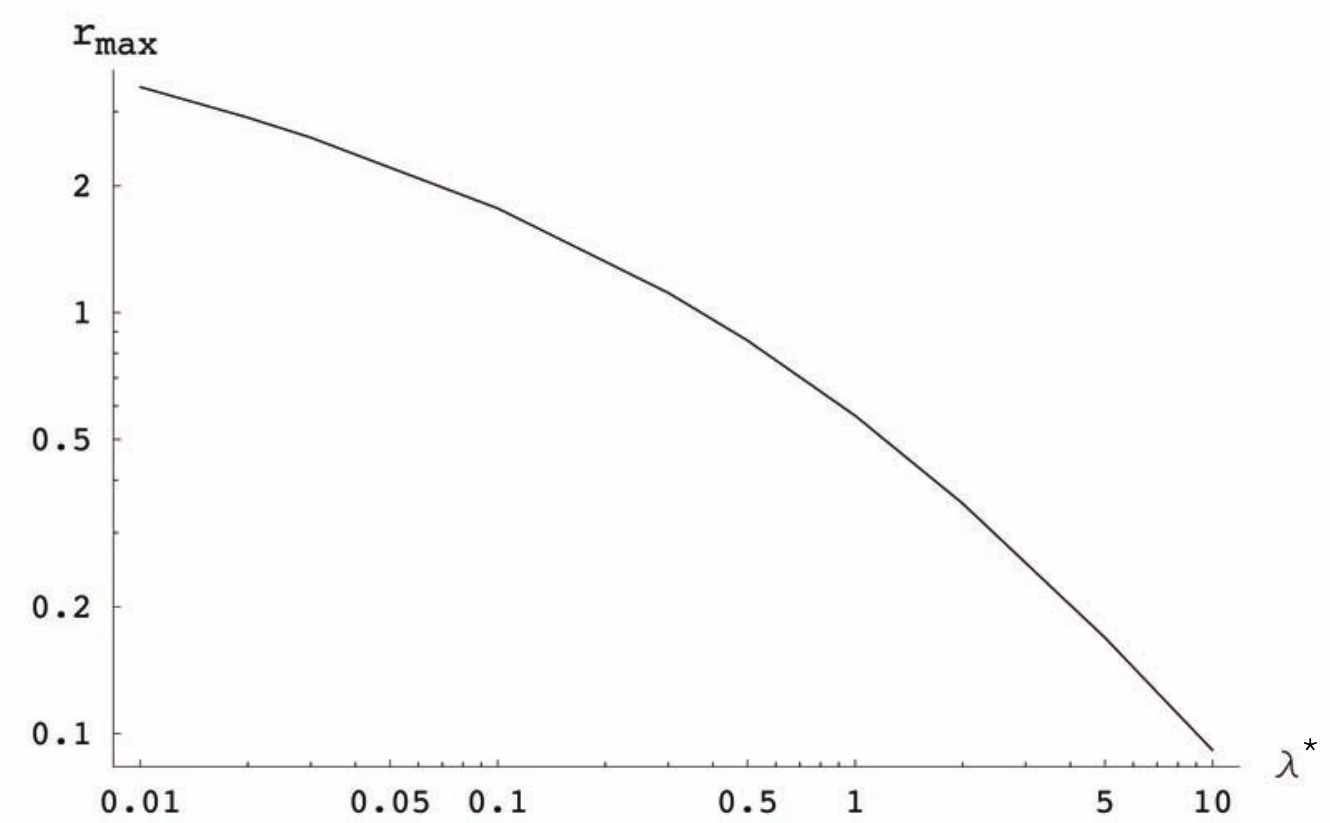}
\caption{A log-log plot of $r_{max}$, in units of $\mu^{-1}$, as a function of $\lambda^*$, measured in units of $\hbar c\mu$, for the massive gluon model.}\label{Yr_maxvslambda}
\end{center}
\end{figure}
At $r_{max}$ one has
\begin{equation}
\left(\frac{dw}{dr}\right)_{r_{max}} =  -\frac{\lambda^*}{\frac{4}{3}g^2} -\mu \left(w+\frac{\lambda^* r}{\frac{4}{3}g^2}\right)_{r_{max}}.\label{Ybc}
\end{equation}
The second derivative is then
\begin{equation}
\frac{d^2w}{dr^2} =  -\beta \frac{w^3}{r^2}+\mu^2(w+\frac{\lambda^* r}{\frac{4}{3}g^2}). \label{wYuk3}
\end{equation}
Notice the extra terms involving $\mu$, which now gives this massless quark equation a scale.

I only consider the the zero pressure case which has $w(r_{max})=0$. The solutions of these equations using Eq.(\ref{Ybc}) as a boundary condition for various values of $\lambda^*$ are illustrated in Fig.~\ref{Ytableau}. Note that in the limit of large $\lambda^*$ the TF functions shrink in size and begin to look suspiciously like the TF $w(r_{max})=0$ function in Fig.~\ref{Rfunctions}. This is not a coincidence. In fact, one can show that Eq.(\ref{wYuk3}) goes over to the simpler Eq.(\ref{diffeq2}) in this limit. On the other hand, for small $\lambda^*$, the TF function grows in size and begins to develop a longer tail. The relationship between $\lambda^*$ and $\beta$ is given in a log-linear plot in Fig.~\ref{Ybetavslambda}. Note that as $\lambda^* \longrightarrow \infty$, we recover the $\beta_f=4.073$ limit for the unconfined massless case considered in the previous section. The relationship between $\lambda^*$ and $r_{max}$ is shown in a log-log plot in Fig.~\ref{Yr_maxvslambda}. From the solutions to Eq.(\ref{wYuk3}) one can show that $r_{max}\longrightarrow 1/\lambda^*$ in the this limit. This is evident from the log-log plot of Fig.~\ref{Yr_maxvslambda}, which gives a slope of -1 in the $\lambda^* \longrightarrow \infty$ limit.

The energy functions can no longer be formally integrated due to the gluon mass term and I will not attempt to study the $A$ dependence of the energy here.

\section{Summary and Comments}

I have introduced a semi-classical TF model of quark interactions in finite quark matter. A MIT like vacuum energy density is used to model confinement, but the (massive) model itself provides effective confinement\footnote{In a quantum treatment, one would expect an exponential tail in the associated wavefunction and this is not true confinement.} from the attractive Coulomb interactions, at least in the absence of CFL locking. I have presented nonrelativistic and relativistic forms of the model and shown that it can also be adapted to the CFL scenario.

I have computed, compiled, and derived a large number of results and relationships, including the TF functions, densities, potentials and other quantities which characterize the various models introduced. However, it must be admitted, in the presentation of topics for investigation I have picked the \lq\lq easy" problems to solve. I have looked at the nonrelativistic case without examining the more difficult coupled equations for unequal quark masses. I have also solved the ultra-relativistic limit after bypassing the much harder massive relativistic equation. Finally, I looked at the CFL modifications for massless quarks without attempting to solve the gap equation for Cooper pairing or incorporating the bag constant. Clearly, one can not make contact with phenomenology without a significant amount of further work on the harder problems. However, the point of this work is  simply to show the range of problems which can be addressed and the relative ease of applicability of the TF model for quarks. Future effort will hopefully connect with the important questions of relative stability and particle phenomenology. Although the TF model assumes a large number of particles, the amazingly accurate binding energies found in the atomic case for low atomic number\cite{spruch} gives us hope that the results obtained from this model will also be reliable for small $A$.

A word about parameters is in order here. I do not expect to fit hadron masses and other measured quantities in order to set the values of parameters in the TF model, which are $\alpha_s$, $B$, and the various quark masses. That is the domain of detailed spectral models. Instead, phenomenologically relevant values of these parameters will be used and explored. Hopefully, many of the relations, for example, hadron mass or binding energy ratios, will be relatively insensitive to these parameters. We will have to see.

Among the many physical applications I have neglected two deserve special mention. One of these is the heavy quark case. In the limit of heavy masses quarks will become nondynamical. This can be modeled by the introduction of central Coulombic color sources, very similar to atomic nuclei, in the midst of dynamical light quark distributions. Their effect on the stability characteristics of the system could be dramatic and such systems are relatively unexplored. In addition, the considerations of Section II can easily be generalized to systems with antiquarks. This will involve introducing new color couplings in the Coulomb interaction between quarks and antiquarks as well as antiquark antiquark interactions in the system energy, $E$. Of course, any possible annihilation diagrams will have to be neglected. Both of these topics will allow a new class of objects to be investigated in this model. Now that the foundations of the model are established, further work can concentrate on fleshing out some of these neglected topics.

\end{document}